\documentclass[12pt,english]{article}

\usepackage{multirow,array}
\usepackage[english]{babel}
\usepackage[utf8x]{inputenc}
\usepackage{amsmath}
\usepackage{bbold}
\usepackage{appendix}
\usepackage{graphicx}
\usepackage[colorinlistoftodos]{todonotes}
\usepackage{setspace}
\usepackage{float}
\usepackage{caption}
\usepackage{subcaption}

\allowdisplaybreaks
\usepackage{ragged2e}

\usepackage{graphicx} 
\graphicspath{ {./images/} }

\usepackage{xcolor}
\usepackage{colortbl}
\usepackage{booktabs}
\usepackage{rotating}
\usepackage{tabularx}
\usepackage{adjustbox}
\usepackage{tablefootnote}
\usepackage{longtable}
\usepackage[flushleft]{threeparttable}

\usepackage[a4paper,bindingoffset=0.2in,%
            left=1in,right=1in,top=1in,bottom=1in,%
            footskip=.25in]{geometry}

\usepackage{stmaryrd}
\usepackage{amsfonts}
\usepackage{titling}

\usepackage{natbib}

\usepackage{setspace}
\usepackage[hang]{footmisc}
\usepackage{lipsum}
\title{Workplace Breastfeeding Legislation and Female Labor Force Participation in the United States}
\author{Julia Hatamyar
    \thanks{ \footnotesize University of York, Centre for Health Economics \newline E-mail: \texttt{julia.hatamyar@york.ac.uk}
    \newline First version: September 13, 2022
    \newline \textbf{Data Availability: }The American Community Survey is publicly available from the US Census Bureau. As per the new conditions of use, data must be obtained through the Panel Study of Income Dynamics public use dataset, produced and distributed by the Survey Research Center, Institute for Social Research, University of Michigan, Ann Arbor, MI. The Infant Feeding Practices Survey II data is only available by request from the CDC. All code used for analysis is available from the author.   
   \newline \textbf{Disclosure Statement:} There was no external funding provided for undertaking this research.}}
\date{\today}

\begin{document}

\begin{titlingpage}
    \maketitle
    \onehalfspacing
    \begin{abstract}
        This paper studies the effects of legislation mandating the provision of workplace breastfeeding amenities on the labor force participation of women in the United States. Using both the American Community Survey and the Panel Study of Income Dynamics, in a staggered difference-in-differences framework, I find evidence that workplace breastfeeding legislation significantly increases the likelihood of female labor force participation (FLFP) across both datasets and multiple specifications, by at least 1.5 percentage points. The timing and magnitude of the post-law increases in FLFP differ across the two datasets. I bolster the analyses using the CDC's Infant Feeding Practices Survey and the Childhood and Adoption Supplement to the PSID, which further suggest an influence of the laws on breastfeeding women. Heterogeneity analysis indicates the presence of substantial treatment effect heterogeneity across subgroups, but the findings are specific to the separate datasets. Across both datasets, the legislation appears to be more effective in states where average pre-law FLFP was comparatively low. I also find evidence of a negative spillover effect, whereby women without children and women with older children may have \textit{reduced} their LFP in response to the legislation.  \\

    \end{abstract}
\end{titlingpage}


\section{Introduction}\label{section:intro}

For decades, economists have studied the challenges that arise from balancing family life with labor market participation \citep{becker2009treatise}, leading to a focus on the effectiveness of ``family-friendly" workplace policies and legislation.  Although the United States lags significantly behind other developed nations in its parental leave and part-time work entitlements \citep{blau2013female, olivetti2016evolution}, there are other areas of policy, such as laws targeting working mothers, that could influence the family-friendliness of workplaces and improve women's labor force participation \citep{olivetti2017economic}. One such policy is state legislation requiring a workplace to provide breastfeeding\footnote{In this paper, I use ``workplace breastfeeding" to mean any act of expressing breast milk while at the place of employment, whether that be physical breastfeeding of the child on the premises or pumping of milk for storage and later use.} amenities. This paper primarily investigates the effect of these breastfeeding laws on female labor force participation.


The acts of working and of breastfeeding are in direct competition for a mother's time \citep{roe1999there}. For women who report breastfeeding in the CDC's Infant Feeding Practices Survey II, the mean number of hours per week spent breastfeeding is 12.01 (standard deviation 8.42), and 16.45 (8.98) hours per week in the first 3 months after childbirth. This suggests a possible opportunity cost of breastfeeding, measurable not only in terms of reduced income from work absences, but in reduction in the rate of acquiring human capital. The worker's decision process also impacts her employer, who loses productivity when she postpones returning to work in order to extend breastfeeding durations. The impact of maternal employment on breastfeeding initiation and duration has been examined at length in the medical literature \citep{kurinij1989does, fein1998effect, abdulwadud2007interventions}. Women who return to full-time employment three months after the birth of a child breastfeed for a shorter durations \citep{lubold2016breastfeeding}, and returning to work less than three months after the birth of a child is negatively associated with both breastfeeding duration and initiation \citep{chatterji2005does}.  The converse of these findings, i.e. the impact of breastfeeding on labor market outcomes for women, is less frequently studied in the United States.

Presumably, prior impact evaluation research has been undertaken with the policy goal of increasing breastfeeding rates and durations, not in order to understand how recent increases in breastfeeding rates might be impacting women's employment decisions. This paper therefore attempts to fill this gap in the literature: as the ``motherhood penalty" persists in the labor market \citep{adda2017career},\footnote{The ``motherhood penalty" is described by labor economists as negative labor market outcomes experienced by mothers, compared to fathers and childless women.} it is imperative for economists to understand whether workplace breastfeeding has a positive or negative impact on female labor force participation (FLFP). Although modern policies which promote breastfeeding as a health-driven goal may significantly and negatively impact female labor supply if women delay return to work in order to breastfeed, it is possible that employer-provided breastfeeding amenities may lessen the burden on working women, mitigating these effects. 

Given the interdependence of the decisions to work and to breastfeed, most studies addressing these issues have employed simultaneous equation models or bivariate probit models to account for the endogenous relationship between the two choices. However, this paper differs by examining the impact of breastfeeding-related workplace \textit{policies} -- specifically, state legislation mandating breastfeeding accommodations -- rather than focusing solely on the decision to breastfeed itself. To my knowledge, this study is the first to evaluate the policy impact of such breastfeeding legislation on FLFP in the United States. The staggered implementation of these laws across different states provides a natural experiment that helps mitigate concerns about endogeneity, particularly when controlling for confounding factors, confirming a lack of pre-existing trends, and accounting for the influence of other state-level policies.\footnote{A detailed discussion of these state policies, including their timing and potential confounding effects, can be found in Section 2.} 

Using the American Community Survey (ACS), the Panel Survey of Income Dynamics (PSID), the Childhood and Adoption Supplement to the PSID, and the Infant Feeding Practices Survey II (IFPS), I examine the impact of passage of workplace breastfeeding legislation on female labor force participation (FLFP). In both traditional two-way fixed effect and interaction-weighted \citep{sun2020estimating} difference-in-differences frameworks, I find evidence suggestive of a significant and positive impact of the passing of the legislation on labor force participation of women with young children. In the ACS data, there is a 1.5 percentage point increase in the probability of FLFP starting 6 years after the law is implemented, with the magnitude of coefficients steadily increasing, suggesting a mild but prolonged effect of the laws on FLFP. For the PSID sample, there is a suggested initial increase of 5.9 percentage points in the probablity of labor force participation in the second year after the law is passed, which decreases and becomes less significant over time. Using the CAH supplment to the PSID confirms the initial impact on the probability of FLFP among women who ever breastfed, with larger magnitudes and longer durations, albeit a much smaller sample. On the whole, the mostly consistent results across datasets and specifications provide evidence of a plausible causal effect of the legislation on the FLFP of mothers. 

Recalling that the treatment, implementation of a workplace breastfeeding law, is exogenous ex-ante to an individual woman's decision to breastfeed (i.e., the state legislation being passed at a particular \textit{time} is not driven by the individual's breastfeeding decision), these results suggest that the law may causally induce women to return to work and also to prolong breastfeeding durations.\footnote{I check the exogeneity assumption of female labor force particpation and the legislation being passed at a particular time in Section 5.6.} Because the PSID does not collect information on breastfeeding practices in the CAH supplement until 2017, I then study the Survey of Infant Feeding Practices II from the CDC, in which mothers reported detailed information about their work and breastfeeding decisions in the year after giving birth from 2005-2007. This allows me to explore the effects of the law on the group of women who do breastfeed. The IFPS results suggest a 2.2 percentage point increase in the liklihood of returning to work after birth in states with the breastfeeding legislation, after controlling for contemporaneous breastfeeding.  


To further explore the results, I perform an array of analyses including examining possible influence of the legislation around the birth of the first child, differential impacts across states with higher vs lower FLFP prior to the legislation being passed, and heterogeneity by child's age, occupation, education, and race/ethnicity. I find evidence of a larger impact of the legislation for black and non-hispanic women in the ACS data, and larger effects for college-educated women in the PSID data. I also find that there is a larger, more immediate impact of the legislation in states where FLFP was in the lowest tercile before the law was passed, compared to those where it was in the highest tercile in both main datasets, suggesting that pre-existing labor market conditions may mediate the impacts of the law.\footnote{There is no evidence of pre-trends, which would suggest passing of the law in response to the lower FLFP rates in these states. I explore this in the Robustness section.}

One additional puzzling finding uncovered during the differential analysis according to child age and placebo tests of the effects of the laws on childless women indicates the presence of possible negative spillover effects on women who are not ex-ante impacted by the law, whereby these women may experience a \textit{negative} impact on their liklihood of FLFP. This may be reflecting increased costs of amenity provision or workload demands being passed onto other women, or an impact on firm behavior whereby \textit{all} women are treated differently in anticipation of the breastfeeding accomodation provision being required. 
Conceptually, workplace breastfeeding legislation reduces the worker's cost of breastfeeding but increases firm costs. Mothers may be induced to breastfeed, thinking they will continue easily after returning to work - and only learning about the cost of breastfeeding when they actually do it (see, for example, \cite{kuziemko2018mommy}). There is therefore possible differential selection into breastfeeding based on the policy change. Importantly, this does not mean that the findings of the paper do not suggest a possible impact of the policies on \textit{both} breastfeeding and labor force participation, especially since the policy appears to impact the entire female labor force. 


Related important work by \cite{hauck2020integrating} studies the impact of the United States breastfeeding legislation considered in this paper  from a health perspective, and finds a 2.3 percentage-point increase of breastfeeding rates in states with the legislation compared to those without legislation between 1990-2011. \cite{mandal2014work} suggest that ``supportive workplaces" are related to higher breastfeeding intensity, for example. \cite{del2012does} find that availability of workplace breastfeeding facilities in the UK increases the probability of a mother returning to work by 4 and 6 months after birth, in addition to increasing the probability of breastfeeding in highly educated mothers. Their bivariate probit model, however, does not evaluate the effects of any \textit{policies} geared towards workplace breastfeeding, but rather the individual firm provisions. \cite{hatsor2019breastfeeding} exploit a safety recall of baby formula in Israel as a natural experiment to show that breastfeeding reduces the number of months in which a new mother works. This paper accounts for more instances of U.S. legislation compared to previous literature, studying the impact in states who pass legislation as recently as 2020. In terms of broader contributions, this paper is likely the first to examine the effects of breastfeeding legislation on FLFP in the United States, and contributes to the literature on family-friendly policies, the child penalty, and the gender gap \citep{bertrand2010dynamics, blau2017gender, kleven2019children}. 

\section{Background}\label{section:background}

Before moving to the empirical analysis, here I briefly outline the evolution of breastfeeding practices and workplace policies in the United States. In section 2.2, I describe the legislation and other maternal workplace policies, and show that the presence of these other policies is not correlated with breastfeeding legislation or other state characteristics which may also impact labor force participation. 

\subsection{Breastfeeding in the United States}

Breastfeeding rates have risen steadily in the United States during the past decades, following recommendation by the WHO that infants be exclusively breastfed until 6 months of age. Figure 1 shows that the rate of any breastfeeding at 6 months of age has risen steadily since 2000. In conjunction with this recommendation, the WHO and UNICEF launched the ``Baby-Friendly Hospital Initiative" in 1991,\footnote{This initiative recommends hospitals provide breastfeeding training and support, and removes availability of free infant formula to new mothers.} which has been found to have a slight positive impact on maternal and infant health outcomes in the UK \citep{fallon2019impact}. By 2012, Baby-Friendly hospitals accounted for 7\% of births in the US \citep{perez2016impact}. It is unclear in the literature whether state workplace breastfeeding legislation arose as a result of WHO recommendations or Baby-Friendly hospital initiatives, and lack of data unfortunately precludes analysis of any effect of these initiatives on the state legislatures.\footnote{See \cite{murtagh2011working}} In the next sections, I discuss the legislation and other possible confounding effects related to the timing of its implementation by state. 

\begin{figure}[H]
\centering
\caption{}
\begin{minipage}{0.95\textwidth} %
\centering
\includegraphics[width=14cm]{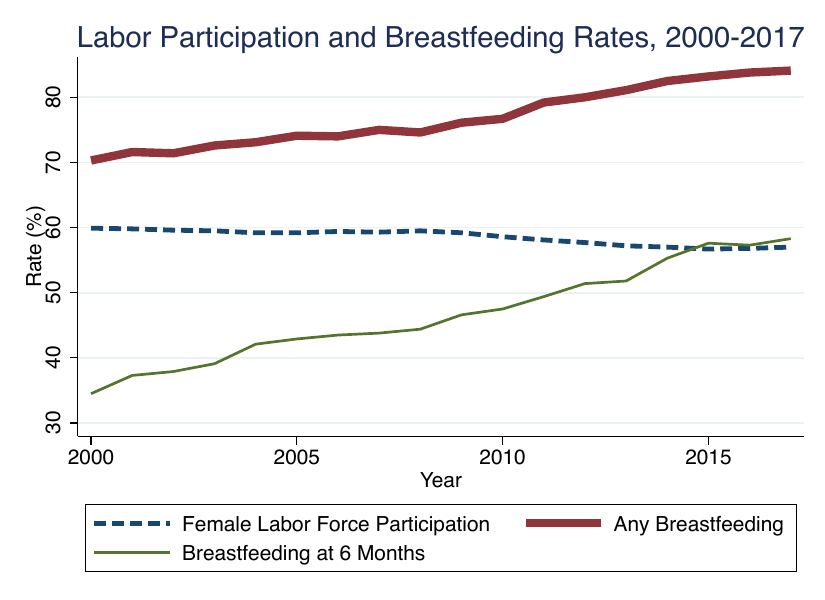}
{\scriptsize \justifying \singlespacing{Data on United States female labor force participation taken from the BLS. Data on breastfeeding rates are from the CDC \textit{National Immunization Survey}, which reports breastfeeding by year of infant birth. Any Breastfeeding indicates whether a child ever was fed breast milk. Breastfeeding at 6 months indicates whether a child was fed breast milk by the age of 6 months, both exclusively and in combination with other food sources.} \par}
\end{minipage}
\end{figure}

\subsection{Workplace Breastfeeding Legislation and Other Maternal Policies}

Individual states began introducing legal statutes requiring or regulating various workplace breastfeeding amenities in 1998 (Table 1). As of this writing, there are 25 states with such statutes.\footnote{I provide a list of statutes used in this paper in the Appendix.} The requirements range from requiring provision of break time for breastfeeding and pumping and requiring access to facilities and locations for such purposes, to requiring employees to be allowed to use their break times to express milk, to protecting female employees from job termination for engaging in lactation activities during breaks. Qualitative research has found that the majority of breastfeeding working mothers do not engage in physical breastfeeding during work hours, instead primarily or always pumping milk at work \citep{felice2017breastfeeding} - however, the legislation often does not distinguish between the two practices. In some states, break time is explicitly provided only for pumping. 

Figure 2 geographically depicts the presence of breastfeeding legislation by US state. It is clear these laws are not clustered in one area, and they are present in different political and cultural climates (California and Arkansas being arguably politically dissimilar, for example). 

\begin{figure}[H]
\centering
\begin{minipage}{0.65\textwidth} 
\caption{Breastfeeding Legislation States}
\includegraphics[width=10cm]{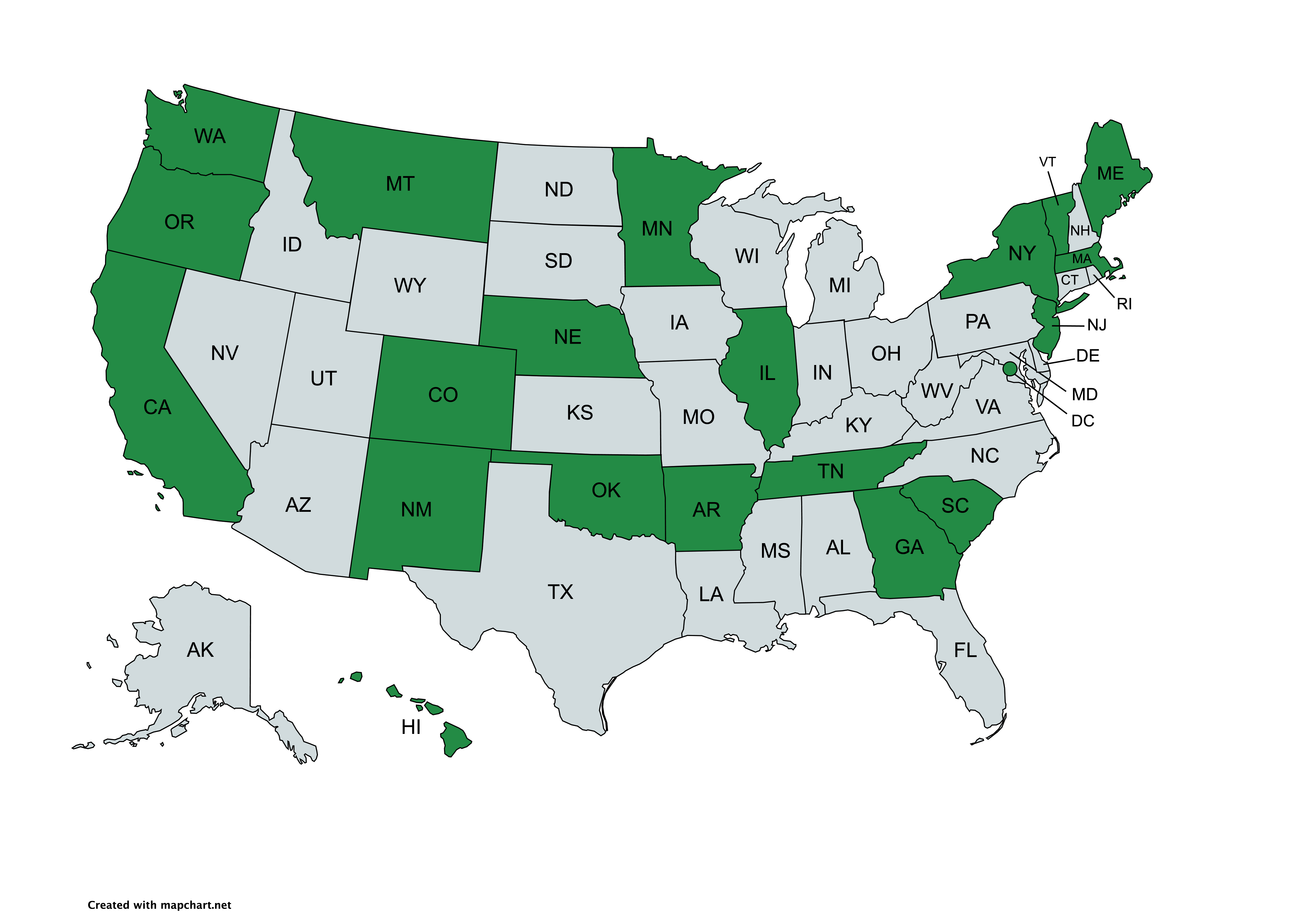}
\end{minipage}
\end{figure}

There may be some concern that workplace breastfeeding legislation was enacted in tandem with other policies which could influence a mother's labor force participation decisions. However, as shown in Table 1, there are only three states which provide paid maternity leave, 15 states with job protection for mothers taking leave after childbirth (whether paid or unpaid), and the timings of these policy implementations never coincide.\footnote{Information on other policies taken from \cite{gault2014paid}.} Table 2 reports correlations between the overall presence of these policies, maternal labor force participation, and average childcare costs in 2013. Importantly, there is no evidence of a relationship between maternal labor force participation and the presence of workplace breastfeeding policies. Although there is mild correlation between the presence of the breastfeeding legislation and job protection following family leave, Table 1 indicates that the timing of these laws is unrelated. There is also no discernible pattern: some states enacted job protection before breastfeeding legislation, and some afterwards, from an identification perspective, the timing of paid leave or job protection laws is unlikely to interfere with any short-term impacts of the breastfeeding legislation.\footnote{To check this more formally, I perform a robustness check which includes a time-varying indicator for the presence of one or both of these other policies. Results are reported in the Appendix.}

\begin{table}[htbp]
  \centering
  \caption{Maternity Policies by State}
  \begin{adjustbox}{width=\textwidth}
  \begin{threeparttable}
    \begin{tabular}{lcccccccc}
    \textbf{State} & Workplace Breastfeeding & & Paid Leave & & Job Protection & & Maternal LFP & Childcare Costs \\
    \hline
    Alabama & No    & -     & No    & -     & No    & -     & 70.1\% & \$5,547 \\
    Alaska & No    & -     & No    & -     & No    & -     & 68.3\% & \$10,280 \\
    Arizona & No    & -     & No    & -     & No    & -     & 59.8\% & \$9,166 \\
    Arkansas & Yes   & 2009  & No    & -     & No    & -     & 62.9\% & \$5,933 \\
    California & Yes   & 2001  & Yes   & 2002  & Yes   & 1993  & 62.1\% & \$11,628 \\
    Colorado & Yes   & 2008  & No    & -     & No    & -     & 65.1\% & \$13,143 \\
    Connecticut & Yes   & 2001  & No    & -     & Yes   & 2013  & 72.2\% & \$13,241 \\
    Delaware & Yes    & 2013     & No    & -     & No    & -     & 70.8\% & \$9,058 \\
    District of Columbia & Yes   & 2007  & No    & -     & Yes   & 2013  & 67.5\% & \$21,948 \\
    Florida & No    & -     & No    & -     & No    & -     & 67.7\% & \$8,376 \\
    Georgia & Yes   & 1999  & No    & -     & No    & -     & 68.5\% & \$7,025 \\
    Hawaii & Yes   & 2013  & No    & -     & Yes   & 1994  & 65.9\% & \$11,748 \\
    Idaho & No    & -     & No    & -     & No    & -     & 58.4\% & \$6,483 \\
    Illinois & Yes   & 2001  & No    & -     & No    & -     & 69.8\% & \$12,568 \\
    Indiana & Weak  & 2008  & No    & -     & No    & -     & 67.6\% & \$8,281 \\
    Iowa  & No    & -     & No    & -     & No    & -     & 77.0\% & \$9,185 \\
    Kansas & No    & -     & No    & -     & No    & -     & 69.9\% & \$10,787 \\
    Kentucky & No    & -     & No    & -     & No    & -     & 66.1\% & \$6,194 \\
    Louisiana & Weak  & 2013  & No    & -     & No    & -     & 68.6\% & \$5,655 \\
    Maine & Yes   & 2009  & No    & -     & Yes   & 2012  & 75.3\% & \$9,360 \\
    Maryland & No    & -     & No    & -     & No    & -     & 72.5\% & \$13,897 \\
    Massachusetts & Yes    & 2018     & No    & -     & Yes   & 2009  & 72.5\% & \$16,549 \\
    Michigan & No    & -     & No    & -     & No    & -     & 68.4\% & \$9,724 \\
    Minnesota & Yes   & 1998  & No    & -     & Yes   & 2014  & 74.8\% & \$13,993 \\
    Mississippi & No    & -     & No    & -     & No    & -     & 73.4\% & \$5,496 \\
    Missouri & \textit{No}*    & -     & No    & -     & No    & -     & 70.4\% & \$8,736 \\
    Montana & Yes   & 2007  & No    & -     & Yes   & 1995  & 69.8\% & \$8,858 \\
    Nebraska & Yes    & 2015     & No    & -     & No    & -     & 72.9\% & \$9,100 \\
    Nevada & No    & -     & No    & -     & No    & -     & 65.9\% & \$10,095 \\
    New Hampshire & No    & -     & No    & -     & Yes   & 1992  & 71.1\% & \$11,901 \\
    New Jersey & Yes   & 2018  & Yes   & 2009  & Yes   & 1989  & 69.3\% & \$11,534 \\
    New Mexico & Yes   & 2007  & No    & -     & No    & -     & 62.2\% & \$7,523 \\
    New York & Yes   & 2007  & Yes & 2021     & No    & -     & 67.5\% & \$14,508 \\
    North Carolina & No    & -     & No    & -     & No    & -     & 70.8\% & \$9,107 \\
    North Dakota & No    & -     & No    & -     & No    & -     & 71.3\% & \$7,871 \\
    Ohio  & No    & -     & No    & -     & No    & -     & 70.6\% & \$7,771 \\
    Oklahoma & Yes   & 2006  & No    & -     & No    & -     & 63.9\% & \$7,741 \\
    Oregon & Yes   & 2007  & No    & -     & Yes   & 2012  & 66.0\% & \$11,078 \\
    Pennsylvania & No    & -     & No    & -     & No    &       & 70.5\% & \$10,470 \\
    Rhode Island & Yes   & 2003  & Yes   & 2013  & Yes   & 1987  & 71.7\% & \$12,662 \\
    South Carolina & Yes    & 2020     & No    & -     & No    & -     & 68.7\% & \$6,372 \\
    South Dakota & No    & -     & No    & -     & No    & -     & 80.3\% & \$5,571 \\
    Tennessee & Yes   & 1999  & No    & -     & Yes   & 2005  & 67.7\% & \$5,857 \\
    Texas & No    & -     & No    & -     & No    & -     & 61.9\% & \$8,619 \\
    Utah  & No    & -     & No    & -     & No    & -     & 52.8\% & \$8,052 \\
    Vermont & Yes   & 2008  & No    & -     & Yes   & 2013  & 69.4\% & \$10,103 \\
    Virginia & Yes   & 2014  & No    & -     & No    & -     & 69.2\% & \$10,028 \\
    Washington & Yes    & 2019     & No    & -     & Yes   & 2010  & 63.0\% & \$12,332 \\
    West Virginia & No    & -     & No    & -     & No    & -     & 61.3\% & \$7,800 \\
    Wisconsin & No    & -     & No    & -     & Yes   & 2011  & 76.6\% & \$11,342 \\
    Wyoming & No    & -     & No    & -     & No    & -     & 65.3\% & \$9,233 \\
    \hline
    \end{tabular}%
    \begin{tablenotes}
            \item[a] This table reports implementation of workplace breastfeeding legislation (in terms of requiring employers to provide space and opportunity to breastfeed or express milk) as well as paid maternal leave policy and leave job protection. 
            \item[b] Paid Leave refers to legislation mandating partial or fully paid leave following the birth of a child. Job Protection legislation mandates the ability to return to work after family or parental leave is taken, whether paid or unpaid. Data on legislation taken from Gault et al (2014). 
            \item[c] Maternal LFP is percentage of mothers in the labor force in 2013. Childcare Costs is the average cost of childcare in 2013. Both are taken from the IWPR's analysis of the American Community Survey Public Use microdata. 
            \item[*] Missouri prevents discrimination against breastfeeding mothers but does not require provision of workplace amenities for breastfeeding and/or pumping.
        \end{tablenotes}
    \end{threeparttable}
    \end{adjustbox}
\end{table}%

\begin{table}[htbp]
  \centering
  \caption{Maternity Policy Correlations in 2013}
  \begin{adjustbox}{width=\textwidth}
  \begin{threeparttable}
\def\sym#1{\ifmmode^{#1}\else\(^{#1}\)\fi}
\begin{tabular}{l*{5}{c}}
\hline\hline
          &\multicolumn{5}{c}{}                                                                          \\
          &    Breastfeeding Law         &    Paid Leave\tnote{b}         &    Job Protection         &  Maternal LFP         & Childcare Costs         \\
\hline
Breastfeeding Law     &     1.00         &                  &                  &                  &                  \\
Paid leave     &     0.29\sym{*}  &     1.00         &                  &                  &                  \\
Job Protection     &     0.43\sym{**} &     0.37\sym{**} &     1.00         &                  &                  \\
Maternal LFP   &    -0.05         &    -0.03         &     0.18         &     1.00         &                  \\
Childcare Costs &     0.24         &     0.17         &     0.51\sym{***}&     0.13         &     1.00         \\
\hline\hline
\multicolumn{6}{l}{\footnotesize \sym{*} \(p<0.05\), \sym{**} \(p<0.01\), \sym{***} \(p<0.001\)}\\
\end{tabular}
    \begin{tablenotes}
            \item[a] This table reports correlations between the presence of different maternity policies and state characteristics in 2013. 
            \item[b] Information on maternal policies, labor force participation, and childcare costs taken from the IWPR. 
        \end{tablenotes}
    \end{threeparttable}
    \end{adjustbox}
\end{table}%

\section{Data}\label{section:data}

For the main analyses, I use the annual American Community Survey (ACS) , which was conducted annually from 2005-2019, and the  biennial Panel Survey of Income Dynamics (PSID), conduced every two years from 1997-2021. The aim is to analyze the effect of the breastfeeding legislation on women for whom it would ostensibly matter most: those with young children. For the ACS data, there is no data indicating the exact age of the children in the household. However, there is an indicator variable for whether the main respondent gave birth within the last 12 months. I therefore keep all observations who answer yes to this question, which effectively reduces the sample to consist entirely of women. I remove women under the age of 15, as they are not officially considered as workforce participants. I also ensure that each woman is studied at the state level of her place of work, to avoid possible confounding by women who reside in one state but work in another. This alleviates any concerns about women working across state lines to take advantage of new breastfeeding legislation. The indicators for passing of breastfeeding legislation are straightforward to merge with the annual ACS data, and are matched by calendar year. The ACS provides detailed occupational information, which I retain, as well as race and ethnicity, educational attainment, and marital/relationship status variables. I create an indicator for labor force participation if the respondent reports any income, wages or working hours $>0 $. The final sample contains N = 340,914 repeated cross-sectional observations, with roughly 23,000 women for each calendar year. Although use of the ACS data precludes use of the individual fixed effects in regression analysis, this large sample size is reassuring. 

The preparation of the PSID dataset is less straightforward. I first collapse individual observations into household units with separate variables by gender. As the PSID is conducted biennially, I count a wave as having been treated by the breastfeeding law if the law passed in the year of the survey or the year prior. Although this results in a reduction of specificity, it helps account for the fact that employers may take time to implement changes required by the law during the months after it is passed \citep{hauck2020integrating}. I remove any households who have relocated states to address endogeneity concerns - if women are relocating states in order to gain access to workplace breastfeeding amenities after passage of the law, results may be biased. All monetary variables are reported in nominal terms and therefore I adjust them using the CPI for 2021, obtained from the Federal Reserve (FRED). As the variable for labor force participation is coded in an ambiguous manner,\footnote{The same number is used to indicate whether someone is currently working or whether they have never worked} I create a dummy variable for either gender indicating labor force participation if wages or labor income are reported to be $>0$ in any two-year wave. After removing all households without children under age 2 present, the final sample contains 53,966 observations with labor market data from 19,267 women. This sample is of course unbalanced, as women may have multiple children and leave/re-enter the sample during the period. For this reason I treat the sample as a repeated cross-section. In order to take advantage of the PSID panel structure, I later perform a secondary analysis where I do not remove households with children over 2 years old, but estimate an interacted panel regression.

\subsection{Descriptive Statistics}

Descriptive statistics by treatment status are presented in Table \ref{tab:sumstats_ACS_PSID}. Noteably, the two dataset samples differ greatly in terms of female characteristics. The ACS sample is younger (29.5 years vs 32 years), more predominantly white, and with a higher proportion of women obtaining a Bachelor's degree or higher (30\% vs 23\%). The PSID sample is poorer, has a large proportion of black respondents (40\% vs 12\%, highly incommensurate with the actual demographic distribution of the United States according to Census data),\footnote{This could be due to attempts during implementation of the PSID to oversample low-income families.} and a higher proportion of women who attended some college but did not obtain a Bachelor level degree. 

\begin{table}[H]\centering
\singlespacing
\def\sym#1{\ifmmode^{#1}\else\(^{#1}\)\fi}
\caption{Summary Statistics by Treatment: ACS vs. PSID Main Samples}
\begin{adjustbox}{width=\textwidth}
\begin{threeparttable}
\begin{tabular}{l c c c | c c c}
\hline\hline
&\multicolumn{3}{c}{\textbf{ACS\tnote{a} }  (2005-2019)}&\multicolumn{3}{c}{\textbf{PSID} (1997-2021)} \\
                &\multicolumn{1}{c}{BF Law\tnote{b}}&\multicolumn{1}{c}{No Law}&\multicolumn{1}{c}{All}&\multicolumn{1}{c}{BF Law}&\multicolumn{1}{c}{No Law}&\multicolumn{1}{c}{All}\\
\hline
FLFP            &     0.62&     0.62&     0.62&     0.70&     0.74&     0.72\\
                &   (0.49)&   (0.48)&   (0.49)&   (0.46)&   (0.44)&   (0.45)\\
                [1em]
Income (\$2021\tnote{c})      & 28,609.59& 24,379.57& 25,794.44& 22,117.97& 22,343.74& 22,246.58\\
                &(47160.04)&(37553.51)&(41066.37)&(31299.63)&(27962.29)&(29444.94)\\
                [1em]
Age         &    30.07&    29.34&    29.58&    33.07&    32.52&    32.75\\
                &   (5.81)&   (5.74)&   (5.78)&   (9.57)&   (9.58)&   (9.58)\\
Single          &     0.22&     0.25&     0.24&     0.26&     0.33&     0.30\\
                &   (0.41)&   (0.43)&   (0.43)&   (0.44)&   (0.47)&   (0.46)\\
Some College         &     0.52&     0.48&     0.49&     0.76&     0.77&     0.77\\
                &   (0.50)&   (0.50)&   (0.50)&   (0.43)&   (0.42)&   (0.42)\\
Bachelor's or more &     0.33&     0.29&     0.30&     0.24&     0.21&     0.23\\
                &   (0.47)&   (0.45)&   (0.46)&   (0.43)&   (0.41)&   (0.42)\\
                [1em]
Black          &     0.10&     0.12&     0.12&     0.32&     0.46&     0.40\\
                &   (0.31)&   (0.33)&   (0.32)&   (0.47)&   (0.50)&   (0.49)\\
Hispanic         &     0.14&     0.16&     0.15&     0.06&     0.02&     0.04\\
                &   (0.34)&   (0.37)&   (0.36)&   (0.23)&   (0.14)&   (0.19)\\
Asian          &     0.08&     0.05&     0.06&     0.02&     0.01&     0.01\\
                &   (0.27)&   (0.21)&   (0.23)&   (0.15)&   (0.09)&   (0.12)\\
                [1em]
Childcare Expense(\$2021) &         &         &         &  1,477.68&  1,379.88&  1,421.97\\
                &         &         &         &(3908.18)&(3476.07)&(3668.57)\\
\hline
Observations    &   114,030&   226,884&   340,914&    23,194&    30,698&    53,892\\           
\hline\hline
\end{tabular}
\begin{tablenotes}
            \item[a] This table reports mean values for summary statistics and standard errors for the main datasets used in the paper. The ACS sample is of women who gave birth during the previous calendar year and runs from 2005-2019. The PSID sample is of women who have children in the home under age 3, and runs biannually from 1997-2021.
            \item[b] ``BF Law" are those who reside in a law enacting state, during the time after the law is passed. 
            \item[c] Household income and childcare costs were adjusted to 2021 levels using FRED data. 
        \end{tablenotes}
    \end{threeparttable}
\label{tab:sumstats_ACS_PSID}
\end{adjustbox}
\end{table}

In both datasets, there are more women with Bachelor degrees and fewer black women in the breastfeeding law enacting states compared to non-law-enacting states. In the ACS data, the average income level in the law enacting states is much higher (\$28,609 vs \$24,379). For the PSID respondents, the average childcare expenses are higher in the law enacting states.

\subsection{Workplace Legislation}

Data on maternity policies including workplace breastfeeding laws, paid leave and job protection laws, is partly taken from \cite{hauck2020integrating} and updated to include more recent years with information from the National Conference of State Legislatures and confirmed by manually checking state government statutes. In three states, breastfeeding statues only apply to public employees, and in two states only to public school employees. Unlike \cite{singh2018integrating} and \cite{hauck2020integrating}, who study the impact of U.S. breastfeeding legislation on breastfeeding rates, I do not include these states in my analysis, since such limited provisions do not impact the entire labor market.\footnote{Due to the data availability in the ACS indicating whether an individual is a state or other local public employee, in the Robustness Section, I repeat the analysis using only those states where the laws apply to public employees as the treated group, and find mixed evidence of effects of these weaker laws.} I also exclude states in which the law only ``encourages" firms to provide lactation support. In sum, for analysis in this paper I define 22 states which introduced their own legislation \textit{requiring} provision of amenities or break-time for \textit{all} workers sometime between 1995 and 2019 as the treated group. 

\section{Methodology}\label{section:methods}

In this section I describe the main empirical models used in the paper, the two-way fixed effects and Interaction-Weighted estimation technique for staggered difference-in-differences (event studies).\footnote{In later sections, I also describe a model for difference-in-differences with an interaction term, and specifications that are not in an event-study framework, which I use for the IFPS II data.}

In an event study framework, regression coefficients are interpreted as the average treatment effect for all cohorts that have experienced \textit{l} periods of treatment (the dynamic ATT). There is a growing literature examining the validity of these difference-in-difference estimates (of which the event study framework is a type) \citep{goodman2018difference, athey2018design, de2020two, callaway2018difference, imai2020use}.  \cite{sun2020estimating}, among others, show that these weighted averages are in actuality not causally interpretable, since they do not necessarily identify convex averages of cohort specific average treatment effects. This is due to estimate weights which do not coincide with actual sample frequencies.\footnote{\cite{borusyak2017revisiting} further discuss the consequences of weighting problems in the estimation of average treatment effects in event studies} Long-run treatment effects are potentially downweighted for cohorts with early treatment onset. 


There are many new proposed methods to address this weighting issue raised in \cite{goodman2018difference}. In this paper, I use the correction proposed by \cite{sun2020estimating} - importantly, this method is designed specifically with time-varying treatment effects in mind, for a staggered treatment timing setting. In contrast, the \cite{de2020two} method is more generally applied to two-way fixed effects settings without being specific to staggered treatment timing,\footnote{This method also requires a 2-period stability in treatment} and the \cite{callaway2018difference} method does not allow for time-varying controls, which are arguably important in this setting. My findings (and findings from a traditional two-way FE model) indicate that effects evolve over time and that the \cite{sun2020estimating} approach is most appropriate.  

When a control group is present in an event study model, unit fixed effects do not cause an identification problem (via collinearity) because they ``pin down" the year fixed effects \citep{borusyak2017revisiting}.\footnote{Without a control group, the effects of the passing of relative and absolute time cannot be disentangled.} In all specifications, I keep observations in never-treated states as a control group. As in \cite{sun2020estimating}, I remove the relative period \textit{l} = -1 from the specification to avoid multicollinearity among the relative period indicators $D^\ell_{i,t}$. The main analysis uses repeated cross-sectional data (to allow comparison across the ACS and PSID), allowing me to include a broad range of individuals across multiple time periods. Importantly, this approach eliminates the need to trim the sample, which could otherwise alter the composition of treated and control groups, a concern raised by Sun and Abraham (2020) regarding dynamic treatment effects. For a secondary analysis, I use the PSID only, in a panel setting and adding in unit fixed effects.

\subsection{Two-way Fixed Effects Estimation}

I first estimate a traditional two-way fixed effects (TWFE) ``event study" regression

\begin{equation}
    Y_{i,t} = \alpha_i + \lambda_t + \sum_{g \in{G}}\mu_g\mathbb{1}\{t - E_i \in g\} + \epsilon_{i,t}
    \label{eq:2wfe}
\end{equation}

on repeated cross sections from the ACS and PSID of $i = 1,...,N$ units for calendar time periods $t= 0,1,...T$. $E_i$ is the initial time of treatment for unit $i$, $Y_{i,t}$ is the outcome of interest for unit $i$ at time $t$, and $\alpha_i$ and $\lambda_t$ are the state and time fixed effects. 

To take advantage of the panel structure of the PSID, I also estimate a TWFE regression with an interaction term for women with a young child of nursing age at the time of the survey:\footnote{Restricting the PSID sample to a balanced panel of women with children of nursing age results in an extremely low number of observations; I use the interaction term in order to retain more individuals.}

\begin{equation}
    Y_{i,t} = \nu_i + \lambda_t + \sum_{g \in{G}}\mu_g\mathbb{1}\{t - E_i \in g\}*\mathbb{1}YoungChild + \epsilon_{i,t}
    \label{eq:int_2wfe}
\end{equation}

where \textit{YoungChild} is a binary indicator for the presence of a child under age 2 or 3, and $\nu_i$ is the \textit{individual} fixed effect. As the sample has not been restricted to only include households with a child of nursing age present, the panel structure is preserved, and the individual fixed effects may be used. 

\subsection{Interaction-Weighted Estimation}

The interaction-weighted estimation process (IWES) used in the main analyses is described below. For further details and proofs, I refer the reader to \cite{sun2020estimating}. Step one of the interaction-weighted estimation process is to use a two-way (state and individual) fixed effects estimation which interacts cohort indicators $\mathbb{1}\{E_i = e\}$ with relative period indicators $D^\ell_{i,t}$: 

\begin{equation}
    Y_{i,t} = \alpha_i + \lambda_t + \sum_{e}\sum_{\ell\neq{-1}}\delta_{e,\ell}(\mathbb{1}\{E_i = e\} \cdot D^\ell_{i,t}) + \epsilon_{i,t}
    \label{eq:iwes1}
\end{equation}

Because there are never-treated cohorts, this estimation is performed on all individuals. Resulting estimates are the $CATT_{e,\ell}$, cohort average treatment effects for each cohort in relative time period $\ell$ to treatment. 

The second step is to estimate weights by sample shares of each cohort in each period $\ell$, i.e. estimating $\Pr\{E_i = e \mid E_i \in{(\ell, T - \ell})\}$.

Lastly, the estimates from the first and second steps are used to form the IW estimator:

\begin{equation}
    \hat{\nu}_g = \frac{1}{|g|}\sum_{\ell \in{g}}\sum_e \hat{\delta}_{e,\ell}\hat{\Pr}\{E_i = e \mid E_i \in{(\ell, T - \ell})\}
    \label{eq:iwes2}
\end{equation}

In other words, the estimator is formed by taking a weighted average of the $CATT_{e,\ell}$ from step 1, $\hat{\delta}_{e,\ell}$, using the weights estimated in step two, the shares of each cohort in the relevant period.

\section{Results}\label{section:results}

In this section, I examine the effect of the enactment of workplace breastfeeding legislation on female labor force participation (FLFP) across a variety of models and datasets. First, I perform repeated cross-sectional regressions (2WFE and IWES) using the ACS and PSID data, focusing on populations who have recently given birth or have young children present in their households. Then, I perform an interacted 2WFE regression using PSID data, where the relative timing of the law is interacted with an indicator for the household having children under age three present. Next, I use the CAH supplement of the PSID and the IFPS data to examine the effects of legislation on FLFP in women who are known to breastfeed. Finally, using the PSID, I examine the effect on FLFP around the timing of the birth of the first child, comparing the effects between those in states with breastfeeding legislation and those without the legislation having been enacted in the two years before childbirth. 

\subsection{Repeated Cross-Section Analysis}

\begin{figure}[H]
    \centering
    \caption{Effect of Legislation on Female Labor Force Participation}
    \begin{subfigure}{.48\textwidth}
        \centering
        \includegraphics[width=\linewidth]{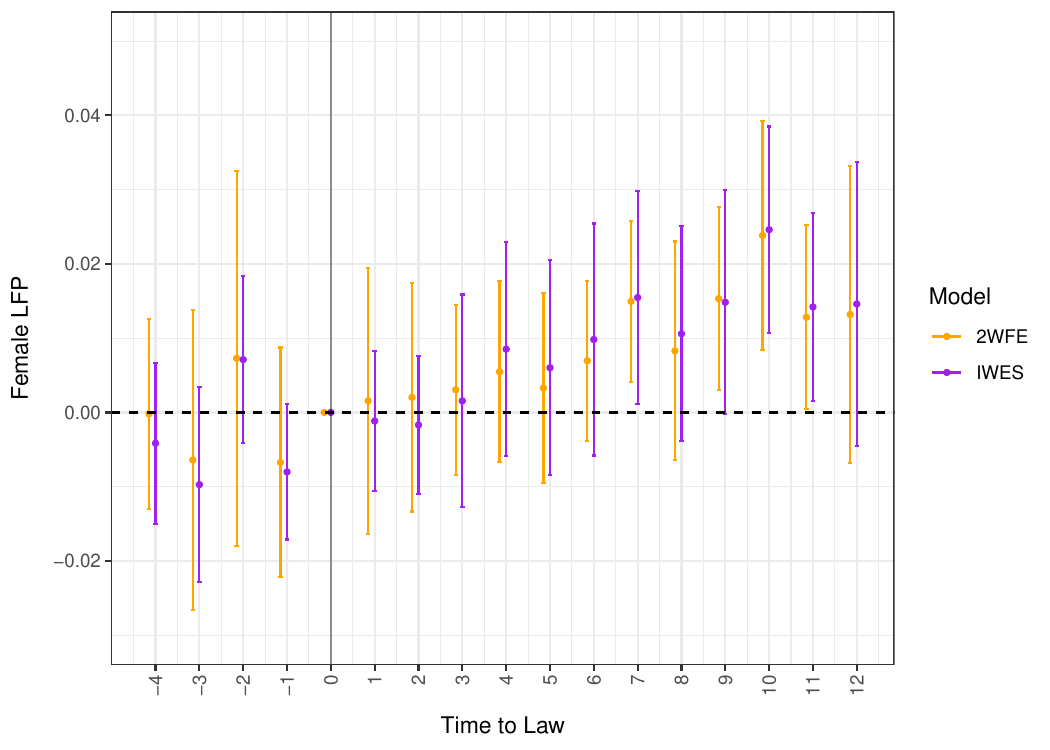} 
        \caption{ACS} 
    \end{subfigure}%
    \hfill
    \begin{subfigure}{.48\textwidth}
        \centering
        \includegraphics[width=\linewidth]{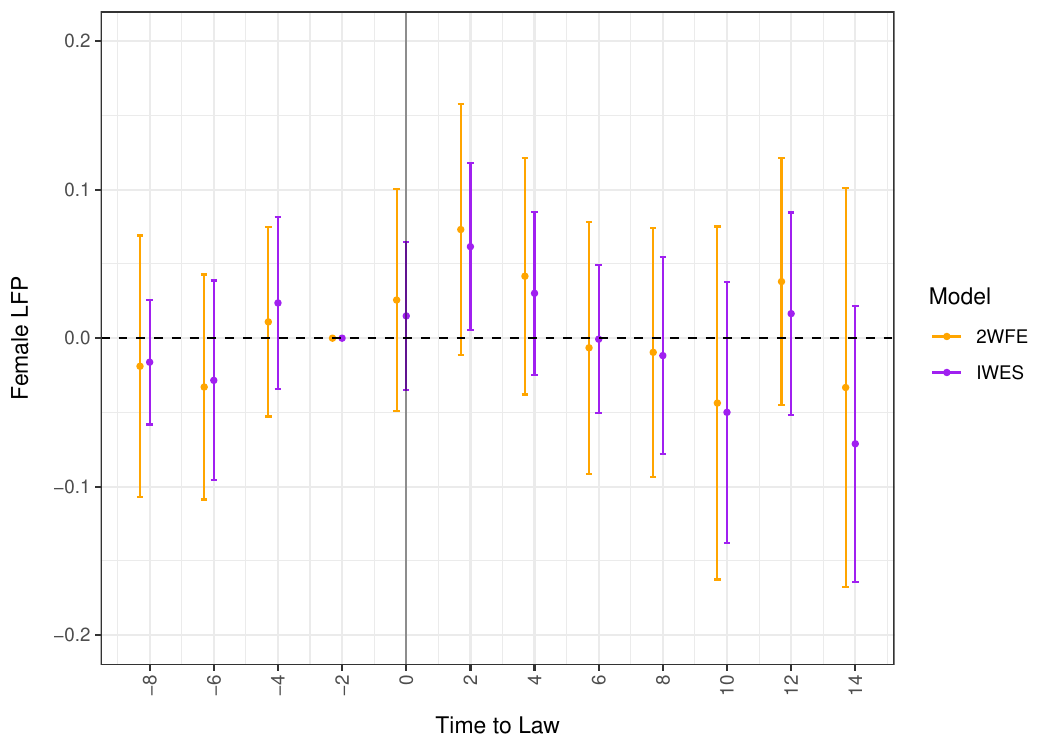} 
        \caption{PSID} 
    \end{subfigure}
    \label{fig:crosssectionLFP}
{\footnotesize \justifying \singlespacing{This figure depicts results from the 2WFE and IWES models (Equations \ref{eq:2wfe} and \ref{eq:iwes2}). 95\% confidence intervals. \textbf{Panel A} shows results using the American Community Survey 2005-2019 sample of women who gave birth in the previous year, and \textbf{Panel B} the PSID from 1997-2021 sample of women with children under age 2. Two-way fixed effect estimates are shown in orange, and IWES estimates are in purple. Time $\ell = -1$ is normalized to zero to account for multicollinearity between relative and calendar time. Point estimates are shown in table \ref{tab:ACS_results} (ACS) Columns 1 and 5, and table \ref{tab:PSID_results} (PSID) columns 1 and 2.} \par}

\end{figure}

Results for the repeated cross-sectional data are depicted in Figure \ref{fig:crosssectionLFP}. Panel A uses the ACS data for women who have given birth in the last year, and shows a persistent increase in female LFP after legislation enactment, which becomes significant after 7 years. According to the IWES model (Table A2 Column 5), the implementation of legislation raises female LFP by around 1.5\% in year 7, and according to the 2WFE estimates, the legislation raises FLFP by the same amount, 1.5\%, in year 8. Panel B however, which uses the unbalanced PSID sample of all women with children under age 3, shows little long-term effect of the law except for an initial significant increase in LFP in the two years after the law is passed. However, the magnitude of the PSID estimates is much larger than the ACS results, with a 5.9\% increase in FLFP in the 2nd year after legislation is enacted according to the IWES results (Table A3, Column 2). 

\subsection{Panel Regression}

I now present results using the PSID but in the interacted model of Equation \ref{eq:int_2wfe} above, where the interaction term is a binary indicator euqal to one if the household contains a child of the younger age group and equal to zero for those with the older age group. This allows the inclusion of multiple observations per individual in the sample, and the use of individual fixed effects. However, the interaction term is not possible to include with the IWES specification; here, I only show results for the TWFE. 

\begin{figure}[H]
    \centering
    \caption{Female Labor Force Participation: Interacted Panel Regressions}
    All Mothers \\
     \vspace{10pt} 
    \begin{subfigure}{.48\textwidth}
        \centering
        \includegraphics[width=\linewidth]{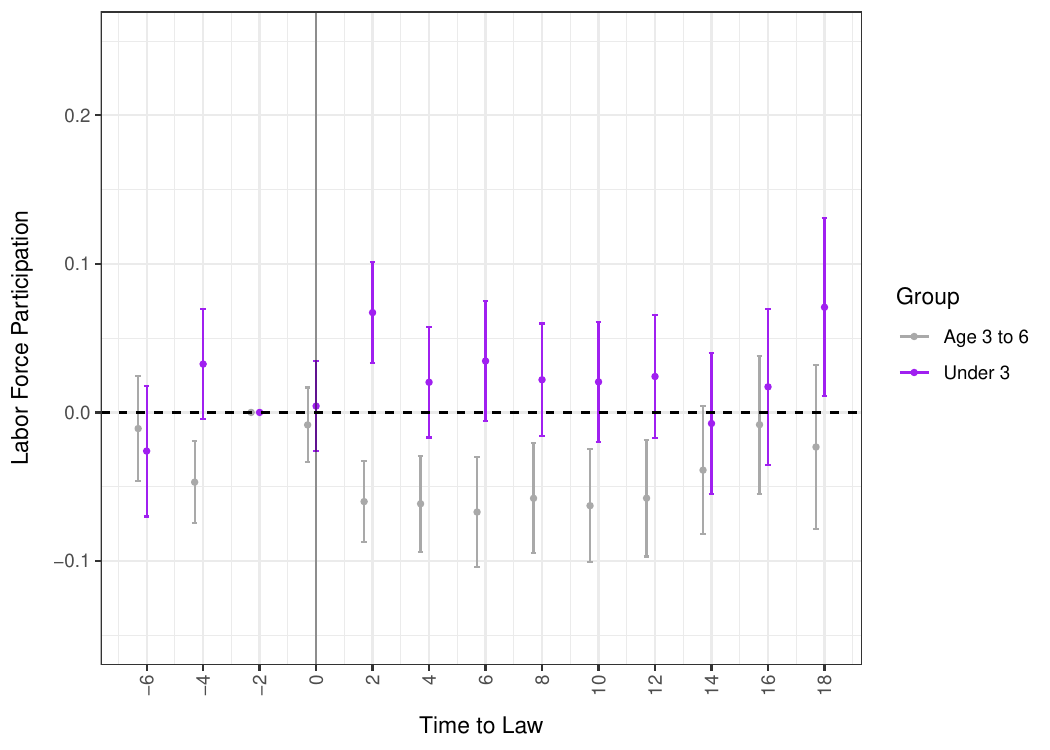} 
        \caption{ }
    \end{subfigure}%
    \hfill
    \begin{subfigure}{.48\textwidth}
        \centering
        \includegraphics[width=\linewidth]{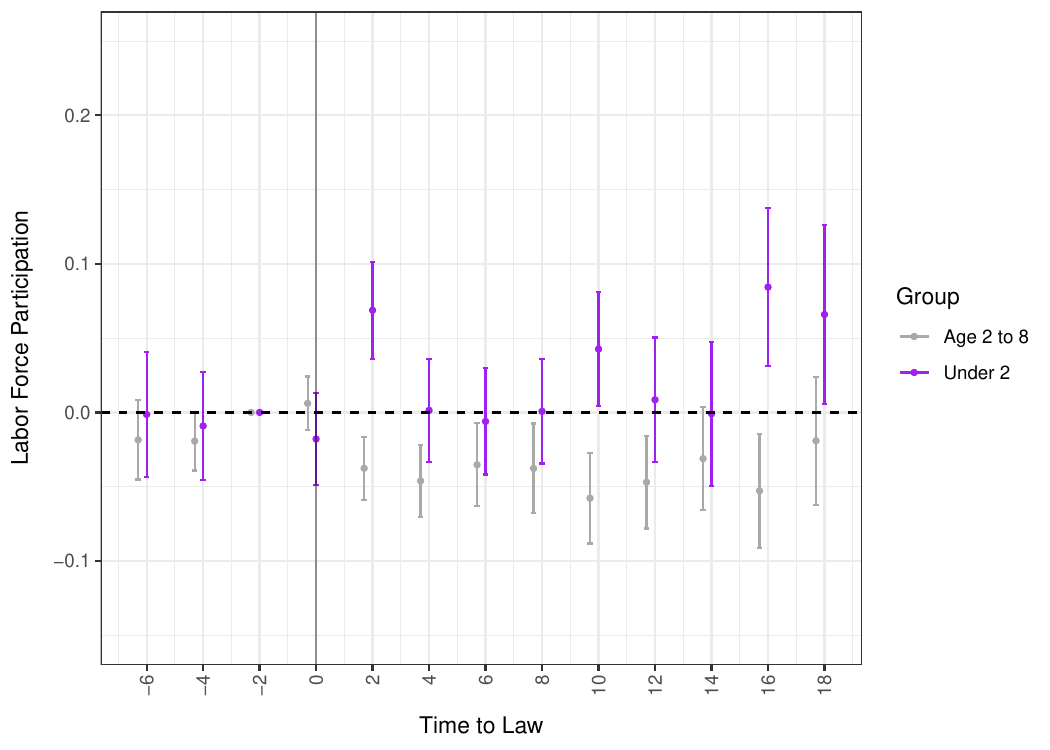} 
        \caption{B}
    \end{subfigure}
    
    \vspace{5pt} 
    
    Mothers with $\leq 3$ Children \\
        \vspace{10pt} 
    \begin{subfigure}{.48\textwidth}
        \centering
        \includegraphics[width=\linewidth]{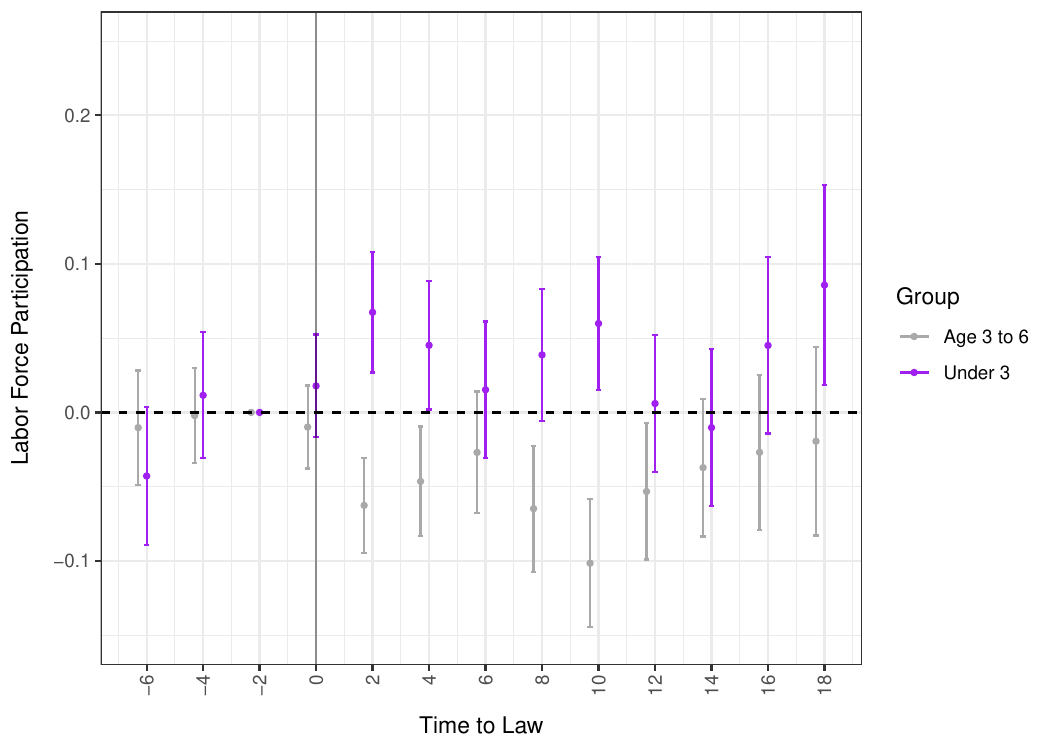} 
        \caption{C} 
    \end{subfigure}%
    \hfill
    \begin{subfigure}{.48\textwidth}
        \centering
        \includegraphics[width=\linewidth]{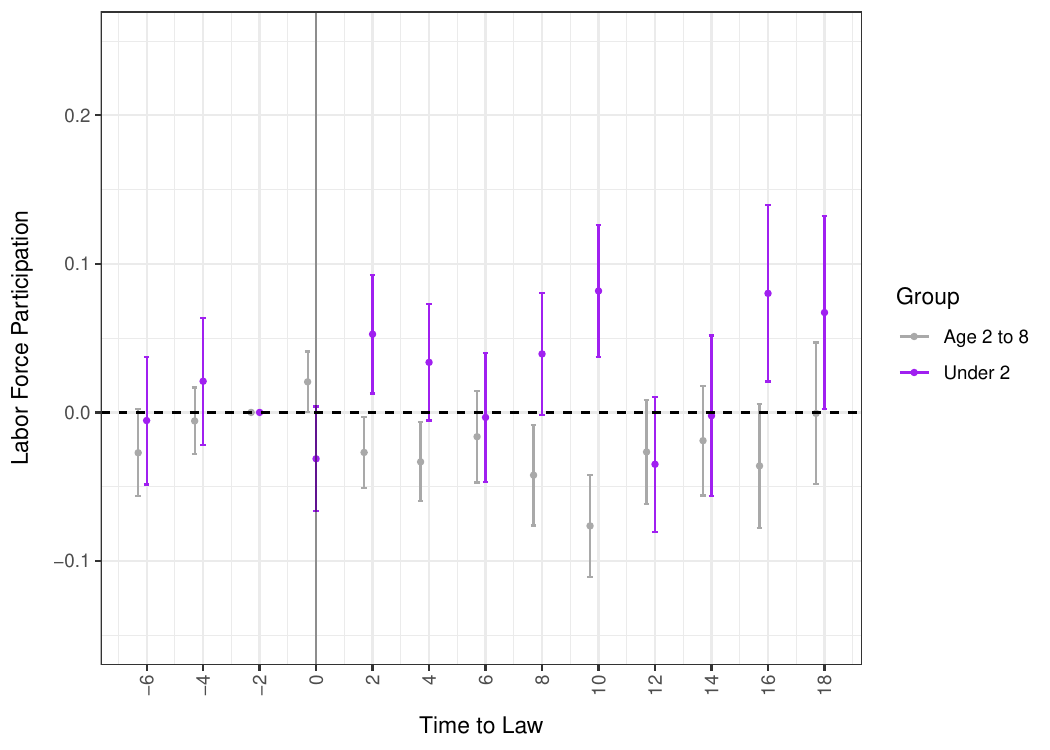} 
        \caption{D} 
    \end{subfigure}
    
    \label{fig:panelLFP}
    
    {\footnotesize \justifying \singlespacing{This figure depicts results from Equation \ref{eq:int_2wfe} with 95\% confidence intervals. Data is taken from the PSID from 1997-2021. The top row shows results comparing those with young children close to nursing age, to slightly older children. Two-way (person and year) fixed effect estimates are shown, with purple for the coefficient on mothers with a child of nursing age. Point estimates and standard errors are reported in the Appendix, Tables \ref{tab:PSID_interactedpanel} and \ref{tab:PSID_interactedpanel_others}. Time $\ell = -1$ is normalized to zero to account for multicollinearity between relative and calendar time.} \par}
\end{figure}

Results for the interacted panel regression are depicted in the top row of Figure \ref{fig:panelLFP}.\footnote{Point estimates and standard errors are reported in the Appendix, Tables \ref{tab:PSID_interactedpanel} and \ref{tab:PSID_interactedpanel_others}. } Panel A compares mothers whose youngest child is under 3 ($YoungChild = 1$) to mothers whose youngest child is between 3 and 6 years old ($YoungChild = 0$), and B (the right side panel) compares mothers whose youngest child is under two ($YoungChild = 1$) to mothers whose youngest child is between 2 and 8 years old ($YoungChild = 0$). Panel A shows a clear difference of the effect of the law for mothers with children of nursing age (under 3) compared to mothers with slightly older children. There is a similar jump in the likelihood of LFP in the first two years after the law begins as was shown in Figure \ref{fig:crosssectionLFP}, however the coefficient remains positive in the subsequent years (although not statistically significant until 12 years after the law enactment). When comparing mothers with children under two to mothers with children between the ages of 2 and 8 (Panel B), the same pattern emerges, but the coefficients for mothers with nursing age children become significant in earlier years. However, in both panels, this analysis has uncovered a puzzling finding: mothers in states with breastfeeding legislation who have young children that are \textit{not} of nursing age appear to be significantly \textit{negatively} impacted by the legislation. Those mothers with children above nursing age (above 3 in Panel A and above 2 in Panel B) experience a significant and persistent drop in the likelihood of LFP for over a decade. I further decompose the sample to explore these interesting results below. 

If mothers with slightly older children have a decreased likelihood of LFP, this may be due to other characteristics of caretaking responsibilities. Mothers whose children are young but not of nursing age may comprise the most balanced control group, as they still face time constraints due to caretaking responsibilities. However, given those responsibilities, those women with many children compared to those with only a few children may not behave similarly in terms of LFP. Therefore I repeat the analysis after restricting the sample to those mothers who have 3 children or less.\footnote{Note that this restriction reduces the length of time for which some mothers remain in the sample and results in a less balanced panel. Results are unchanged if mothers who \textit{ever} have more than 3 children are removed from the entire sample, which is more balanced but contains fewer observations.} Results are shown in the bottom row of Figure \ref{fig:panelLFP}, panels (C) and (D).\footnote{These results are consistent with further tests (See Robustness Section) which indicate a possible spillover effect of the legislation on childless women. I discuss these implications in later sections.} I further explore these puzzling findings in the Robustness Section.

\subsection{Women Who Breastfeed}

To explore the effects of legislation using data including information about breastfeeding activity, I now introduce two additional datasets: the Childhood and Adoption Supplement to the PSID, and the Infant Feeding Practices Survey II. 

The PSID dataset is supplemented with the Childhood and Adoption History Supplement from the 2017 PSID (CAH).  This cross-sectional supplement indicates, for each mother and child pair, whether breastfeeding was initiated. I merge these indicators with the main PSID sample by individual in order to analyze the effect of the workplace laws only on women who have indicated that they breastfed a child at some point within the sample period. This allows for interpretation of effect of the law on an ``intention to breastfeed" sample. Not every individual mother in the entire PSID is included in the CAH supplement, and the sample size is significantly reduced when restricting the analysis to this group of households.

In addition to the CAH supplement, I analyze the effect of workplace breastfeeding legislation on mothers who participated in the Infant Feeding Practices Survey II (IFPS), which was conducted by the Food and Drug Administration (FDA) and Centers for Disease Control and Prevention (CDC) in 2005–2007. The IFPS was a longitudinal study following mother-infant pairs from the third trimester of pregnancy through the first year of the child's life. Questions on the survey include detailed information about duration and intensity of breastfeeding, work hours and employment status, as well as a number of questions regarding ease of workplace breastfeeding and pumping. The CDC clearly notes that the IFPS is not taken from a representative sample of the population, which I confirm by comparing characteristics of interest for the IFPS and PSID samples used in this paper. Another drawback of the IFPS data is the fact that individual labor income is not reported, either in aggregate or by period, and only categorical information about household-level income is available. As in previous literature examining post-birth labor market activity, the IFPS II sample is restricted to women who worked before giving birth in order to condition on pre-birth characteristics \citep{mandal2014work}.

\begin{table}[H]
  \centering
  \singlespacing
  \caption{Summary Statistics, by Breastfeeding Status}
  \begin{adjustbox}{width=0.9\textwidth}
  \begin{threeparttable}
    \begin{tabular}{lrcccccc|cccccc}
          &       &       &       & \textbf{PSID (CAH)} &       &       & \multicolumn{1}{c}{} &       &       &       & \textbf{IFPS II} &       &  \\
          &       & Ever BF\tnote{b} &       & Never BF &       & All   &       &       & Ever BF &       & Never BF &       & All \\
    \midrule
    \midrule
   HH Income\tnote{c} (\$) &       & 31,826 &       & 37,084 &       & 36,228 &       &       & 52,267 &       & 46,455 &       & 51,475 \\
          &       & (30,225) &       & (57,831) &       & (54,337) &       &       & (27,149) &       & (28,100) &       & (27,346) \\
    Labor Income\tnote{d} (\$) &       & 11,074 &       & 11,814 &       & 11,693 &       &       & -     &       & -     &       & - \\
          &       & (13,066) &       & (13,528) &       & (13,454) &       &       &       &       &       &       &  \\
   Income Contribute\tnote{d} &       & -     &       & -     &       & -     &       &       & 28.25\% &       & 28.33\% &       & 28.26\% \\
          &       &       &       &       &       &       &       &       & (26.56\%) &       & (26.19\%) &       & (26.50\%) \\
    LFP  &       & 80.21\% &       & 78.52\% &       & 78.79\% &       &       & 52.34\% &       & 52.32\% &       & 52.34\% \\
          &       & (39.88\%) &       & (41.07\%) &       & (40.88\%) &       &       & (49.96\%) &       & (50.02\%) &       & (49.96\%) \\
    Work Hours\tnote{e}  &       & 28.27 &       & 27.08 &       & 27.27 &       &       & 10.25 &       & 12.63 &       & 10.58 \\
          &       & (18.57) &       & (18.53) &       & (18.53) &       &       & (15.15) &       & (16.88) &       & (15.41) \\
    Age   &       & 41.03 &       & 41.52 &       & 41.44 &       &       & 29.15 &       & 29.11 &       & 29.14 \\
          &       & (10.48) &       & (11.04) &       & -10.94 &       &       & (5.43) &       & (5.45) &       & (5.43) \\
    Married &       & 63.25\% &       & 66.55\% &       & 66.02\% &       &       & 77.69\% &       & 69.97\% &       & 76.63\% \\
          &       & (48.25\%) &       & (47.19\%) &       & (47.24\%) &       &       & (41.65\%) &       & (45.91\%) &       & (42.32\%) \\
    White &       & 54.24\% &       & 56.19\% &       & 55.87\% &       &       & 84.81\% &       & 89.47\% &       & 85.45\% \\
          &       & (49.86\%) &       & (49.62\%) &       & (49.66\%) &       &       & (35.90\%) &       & (30.74\%) &       & (35.27\%) \\
    Black &       & 38.34\% &       & 37.84\% &       & 37.92\% &       &       & 4.05\% &       & 4.95\% &       & 4.17\% \\
          &       & (48.66\%) &       & (48.51\%) &       & (48.52) &       &       & (19.72\%) &       & (21.73\%) &       & (20.01\%) \\
  Education &       & 12.54 &       & 12.98 &       & 12.9  &       &       & -     &       & -     &       & - \\
          &       & (2.48) &       & (2.47) &       & (2.48) &       &       &       &       &       &       &  \\
    Some College &       &       &       &       &       &       &       &       & 78.17\% &       & 59.44\% &       & 75.62\% \\
          &       &       &       &       &       &       &       &       & (41.32\%) &       & (49.18\%) &       & (42.95\%) \\
          &       &       &       &       &       &       &       &       &       &       &       &       &  \\
    N     &       & 566   &       & 2,915 &       & 3,481 &       &       & 2,048 &       & 323   &       & 2,371 \\
    \bottomrule
    \end{tabular}%

    \begin{tablenotes}
            \item[a] This table reports mean values for summary statistics and standard errors for the main datasets used in the paper. The 2017 Childhood and Adoption Supplement was merged with the PSID by person ID and collapsed to household units identified in both samples containing at least one adult woman. The resulting PSID sample wave in 2005 is then compared to the first wave of the IFPSII 2005 sample.
            \item[b] ``Ever BF" are those who report breastfeeding in the 2017 CAH supplement to the PSID. For the IFPSII, they are women who ever breastfed during the 12-month post-partum period (in 2005). 
            \item[c] Household income is reported in non-uniform buckets in the IFPSII. Median values were imputed and used to calculate the statistics. 
            \item[d] Individual labor income is not reported in the IFPSII. Rather, female income is reported as a percentage of contribution to total household income. Likewise, number of years of education is not reported in the IFPSII. 
            \item[e] Work hours are reported on an annual basis in the PSID, this is divided by 52 weeks to obtain a weekly average. 
        \end{tablenotes}
    \end{threeparttable}
    \end{adjustbox}
  \label{tab:sumstats_BFonly}%
\end{table}%

Table \ref{tab:sumstats_BFonly} shows comparison of descriptive statistics for each additional dataset, by breastfeeding status, using the PSID wave from 2005 merged with the 2017 CAH supplement to identify households in which a woman ever breastfed. Women in the IFPS II survey are much less likely to work than women in the PSID (52.34\% vs 78.79\%), are younger (29.14 vs 41.44), wealthier (average household income of \$51,475 vs 36,228), more likely to be married (76.63\% vs 66.02\%), and are predominantly white (85.45\% vs 55.87\%). In the PSID sample, women who ever breastfed have higher rates of labor force participation (80.21\%) and work more hours (28.27) than those who do not (78.52\% and 27.08). In contrast, women who breastfeed in the IFPS II sample work fewer hours (10.25 vs 12.63) and have no differences in their rates of labor force participation. Women who report ever breastfeeding have fewer years of education than those who breastfed in the PSID, which contrasts with the fact that breastfeeding rates are increasing in education \citep{dubois2003social}. This is possibly due to the increased age of women in the PSID sample.
\textbf{Results: PSID Childhood and Adoption Supplement} 
\newline 

I report results for the small sample of women who are known to have breastfed (as reported in the CAH supplement to the PSID) in Table 5. For this data, I am able to use both the 2WFE and IWES models. As the CAH supplement was only introduced in 2013, the sample period is shorter and runs for only 5 waves from 2013 - 2021. This means that we are unable to examine long term effects of the legislation. Nevertheless, the results are weakly suggestive of the legislation impacting breastfeeding mothers' LFP in the years immediately following law implementation. 

{\singlespacing
\def\sym#1{\ifmmode^{#1}\else\(^{#1}\)\fi}
\begin{table}[H]
\caption{FLFP: Mothers who Ever Breastfed (CAH Supplement)}
  \begin{adjustbox}{width=0.95\textwidth}
\begin{tabular}{l c c c c | c c}
\hline\hline
            & \multicolumn{4}{c}{Child $\leq$ 3} & \multicolumn{2}{c}{Law $t$ *Child $\leq$ 3} \\
            &\multicolumn{1}{c}{(1)}&\multicolumn{1}{c}{(2)}&\multicolumn{1}{c}{(3)}&\multicolumn{1}{c}{(4)}&\multicolumn{1}{c}{(5)}&\multicolumn{1}{c}{(6)}\\
            &\multicolumn{1}{c}{2WFE}&\multicolumn{1}{c}{IWES}&\multicolumn{1}{c}{2WFE}&\multicolumn{1}{c}{IWES}&\multicolumn{1}{c}{2WFE}&\multicolumn{1}{c}{2WFE}\\
\hline
$\ell$ = -6         &     -0.0241         &       -.0307           &      -0.0363         &       0.0173              &   0.0510         &      0.0741          \\
            &    (0.0633)         &      (0.0556)               &    (0.107)        &     (0.0698)               &      (0.193)         &     (0.117)             \\
$\ell$ = -4        &       0.103\sym{***}&       0.1169\sym{***}        &      0.104        &        0.0814             &          0.186\sym{***}&       0.158\sym{*}              \\
            &    (0.0291)         &     (0.0242)              &    (0.119)          &          (0.0564)           &              (0.0366)         &    (0.0829)                       \\
$\ell$ = 0          &      0.0972         &       0.1038\sym{**}             &      0.168\sym{*}        &         -0.0037            &        0.205\sym{***}&       0.135\sym{*}               \\            &    (0.0584)         &        (0.0462)         &    (0.1000)         &        (0.0516)             &              (0.0544)         &    (0.0734)               \\
$\ell$ = 2          &       0.107\sym{***}&        0.1086\sym{***}             &      0.105          &         -0.0331             &           0.0406         &     -0.0496                 \\            &    (0.0364)         &   (0.0281)            &    (0.0833)         &         (0.0496)            &           (0.0399)         &    (0.0645)                  \\
$\ell$ = 4         &       0.131         &           0.1609\sym{***}          &     0.0925         &       -0.1663              &            0.103         &     0.00385              \\            &    (0.0800)         &        (.0500)            &    (0.107)         &         (0.1126)            &              (0.0837)         &    (0.0680)            \\

$\ell$ = 6          &      -0.255         &                     &       0.222\sym{**} &                     &                     &                     \\
            &     (0.152)         &                     &     (0.113)         &                     &                     &                     \\

Unmarried      &       0.0410         &      0.0417         &      0.0509         &      0.0597         &      0.0759\sym{**} &    -0.00424  \\
            &    (0.0404)         &    (0.0404)         &    (0.0640)         &    (0.0436)         &    (0.0286)         &    (0.0303)    \\
 & & & & & & \\

Constant      &      0.402\sym{**} &       0.387\sym{***}&       1.321\sym{***}&       1.239\sym{***}&       0.434\sym{***}&       1.146\sym{***}\\            &     (0.147)         &     (0.141)         &     (0.334)         &     (0.233)         &    (0.0887)         &     (0.245)          \\
\hline
Fixed Effects & State & State & Person & Person & State & Person \\
R-squared   &       0.146         &       0.147         &       0.773         &       0.720         &       0.144         &       0.656      \\
N           &        2662         &        2662         &        2662         &        2130         &        4275         &        4275          \\
Clusters    &           31         &          31         &        1373         &         841         &          31         &        1352       \\
\hline\hline
\multicolumn{7}{l}{\footnotesize Standard errors in parentheses. All models include year fixed-effects.}\\
\multicolumn{7}{l}{\footnotesize \sym{*} \(p<0.1\), \sym{**} \(p<0.05\), \sym{***} \(p<0.01\). Control coefficients reported in Appendix.} \\
\end{tabular}
\end{adjustbox}
\label{tab:CAHmain}
\end{table}
}

Column (2) in Table \ref{tab:CAHmain} shows a persistent 10\% increase in LFP for mothers with children under the age of three in the 6 years following legislation start, using the IWES model. However, there is evidence of a positive pre-trend in this sample. When using person fixed effects, the results are no longer significant for the IWES model (and the pre-trends dissapear), but show a slight increase in the 2WFE model for the immediate year following implementation of the law, and later years.\footnote{Due to small sample size, it is not possible to run the interacted regression using event time $e = 6$.} In columns (5) and (6), the timing indicator is interacted with a binary indicator for whether there is a child under age 3 in the household (with all households containing at least one child). For example, the first row of column (5) depicts coefficients where the timing indicator for 6 periods before treatment is interacted with the binary indicator for the presence of the child under 3 (see Appendix table \ref{tab:cah_full} for coefficients where the child indicator = 0). There is an immediate effect of the law in its initial year for mothers with younger children compared to mothers with older children whether state or person fixed effects are used, however, the effect does not persist. In many of the specifications, there is evidence of a significant increase in mothers' LFP in the 4 years prior to the law implementation. Whether this represents a violation of parallel trends due to mothers perhaps choosing whether to work in anticipation of the law and also of breastfeeding, is explored in Section \ref{subsection:birthevent}, which examines the effect of the event of childbirth in states with and without the legisltion enacted.

\textbf{Results: IFPS II Sample}
\newline
Because there is no exact information indicating whether mothers in the PSID sample were breastfeeding during the years of law implementation, I now extend the analysis using the IFPS survey data. Results are reported in Table \ref{tab:IFPS_results}. The IFPS II was conducted from 2005-2006, so I include a state as having been treated by the law if the law went into effect any year prior to 2006. I first examine the impact of having breastfed on the likelihood of labor force participation, then add the law indicator, then finally the interaction term. The interacted variable of interest, $WorkplaceLaw*Breastfeeding_{i,t}$ represents a dummy for women $i$ who reported breastfeeding during the previous month interacted with the law treatment dummy.\footnote{Although the indicator for breastfeeding is for the previous month and not the previous wave $t$, this is to be consistent with the main outcome variable which only measures LFP during the last calendar month.} As problems with breastfeeding are a significant negative determinant of both work initiation and intensity \citep{mandal2014work}, I include this response as a control variable in all specifications (it should be noted, however, that 88\% of women in the survey report experiencing problems with breastfeeding during the first two weeks after giving birth). Because maternity leave characteristics have been found to influence both work and breastfeeding decisions 
\citep{guendelman2009juggling, baker2008does}, I also control for whether the mother's maternity leave is fully paid, partially paid, or non-paid. 

For analysis of the Infant Feeding Practices Survey II data, I estimate a difference-in-difference regression as in \cite{mandal2014work}. A regression of the following form is estimated:

\begin{equation}
\begin{split}
    Y_{i,t}= & \beta_1(WorkplaceLaw*Breastfeeds_{i,t}) +  \beta_2WorkplaceLaw + \\ 
    & \beta_3*Breastfeeds_{i,t} + X_i + Z_{i,t} + \alpha_i + \lambda_t + \epsilon_{i,t}
    \label{eq:plaindid}
\end{split}
\end{equation}

Where $WorkplaceLaw$ is a dummy indicating residing in a state with a workplace breastfeeding requirement law, $Breastfeeds$ is a dummy indicator for individual $i$ breastfeeding her baby in time $t$, $X_i$ is a vector of time-invariant controls, $Z_{i,t}$ is a vector of time-varying controls, $\alpha_i$ is state fixed-effect and $\lambda_t$ is time fixed-effects. Time fixed-effects are used in order to control for feeding requirements changes a baby experiences over time. Unit fixed effects cannot be used since by definition, the treatment effect remains constant for an individual over time in this sample. 

Importantly, the IFPSII is not a representative sample of the population, and because of potential endogeneity between the treatments, the decision to breastfeed and the presence of the law, the estimates likely cannot not be interpreted causally. Rather, I include this analysis to explore the mechanisms behind the event study specifications. Results are depicted in Table \ref{tab:IFPS_results}.

{\singlespacing
\def\sym#1{\ifmmode^{#1}\else\(^{#1}\)\fi}
\begin{table}[H]
  \centering
  \caption{IFPS II: Labor Force Participation}
  \begin{adjustbox}{width=\textwidth}
  \begin{threeparttable}
    \begin{tabular}{l ccc| cc}
    \toprule
              & (1)   & (2)   & (3)   & (4)   & (5)   \\
              &    &    &    & Law States   & No Law States   \\
    \midrule
    \midrule
Breastfed     &    -0.00509\sym{**} &    -0.00509\sym{**} &    -0.00427\sym{**}                 &    -0.00178         &    -0.00813         \\
            &   (0.00252)         &   (0.00252)         &      (0.00187)                &   (0.00184)         &   (0.00603)         \\
[1em]
Law       &                     &      0.0217\sym{***}&       0.0216\sym{***}              &                     &                     \\
            &                     &   (0.00283)         &      (0.00318)               &                     &                     \\
[1em]
Breastfed*Law &                     &                     &    0.000560         &                     &                     \\
            &                     &                     &   (0.00346)         &                     &                     \\
[1em]

\hline 
Observations           &        2506         &        2506         &        2506         &         622         &        2395         \\
Clusters     &      48         &      48         &      48         &     197         &     740         \\
\hline 
Fixed Effects & State & State & State & Person & Person \\
    \bottomrule
    \end{tabular}%
    \begin{tablenotes}
            \item[a] Results reported for the IFLS II sample, 2005-2007. The outcome variable is working for pay during the previous month. ``Breastfed" indicates whether the respondent had breastfed the child during the previous month. 
            \item[b] Controls in all specifications include education level, age, household income, occupation, race/ethnicity indicators, whether the respondent indicated ever having problems breastfeeding, and whether the respondent had paid maternity leave.
            \item[c] *** p$<$0.01, ** p$<$0.05, * p$<$0.1.
            \item[d] Standard errors are clustered at the state level in columns 1-4, and at the person level in columns 4-5.
        \end{tablenotes}
    \end{threeparttable}
    \end{adjustbox}
  \label{tab:IFPS_results}%
\end{table}%
}

Column (1) in Table \ref{tab:IFPS_results} shows the coefficient in row 1 for regression the labor force participation indicator on breastfeeding, without accounting for the breastfeeing law. There is a significant decrease of about 0.5 percentage points in the likelihood of working, given being recently breastfeeding. Column (2) adds the breastfeeding law indicator, which has a strongly significant coefficient indicating a 2.2 percentage point increase in the probability of working after birth, controlling for breastfeeding status. Column (3) contains the interacted term coefficient, which indicates the effect on women who reside in the law-enacting state who also breastfeed. This coefficient is positive, but very small in magnitude and not statistically significant. Finally, in columns (4) and (5) I examine the impact of breastfeeding on the labor force participation outcome alone in a model incorporating person-specific fixed-effects. Given that there are no other time-varying covariates in the model, it is not possible to examine the impact of the law using person fixed-effects. In this specification there is no significant impact of breastfeeding on the decision to return to work. Howwver, the magnitude of the estimate decreases in states without the breastfeeding law. Taken together, these results are weakly suggestive of an impact of the legislation on maternal FLFP, although the small sample size warrants cautious, likely non-causal interpretation. At the very least, the analysis uncovers possible associations in this novel data between the legislation and the decision to work.

\subsection{Effect Around Timing of Childbirth}\label{subsection:birthevent}

In this section, I implement a variation in the analysis technique to explore whether there is a differential effect of the laws conditional on the age of the child. We may wish to know whether there is a pronounced impact of the law in the years closely after birth compared to the pre-birth years, in states where the breastfeeding laws have been implemented. To examine this, I implement the procedure as in \cite{cortes2023children}, who study the differential effects by gender around the timing of the birth of the first child using PSID data. Specifically, I run the regression as in Equation \ref{eq:int_2wfe}, but the treatment is defined as the birth of the first child ($E_i$ is the time of birth), so the relative time indicators $\ell = t - E_i$ indicate the event-time years before and after birth. As in \cite{cortes2023children} I run the analysis for various groups, with the interaction term defined accordingly.

\begin{figure}[H]
    \centering
    \caption{Labor Force Participation around Birth of First Child}
    \begin{subfigure}{.48\textwidth}
        \centering
        \includegraphics[width=\linewidth]{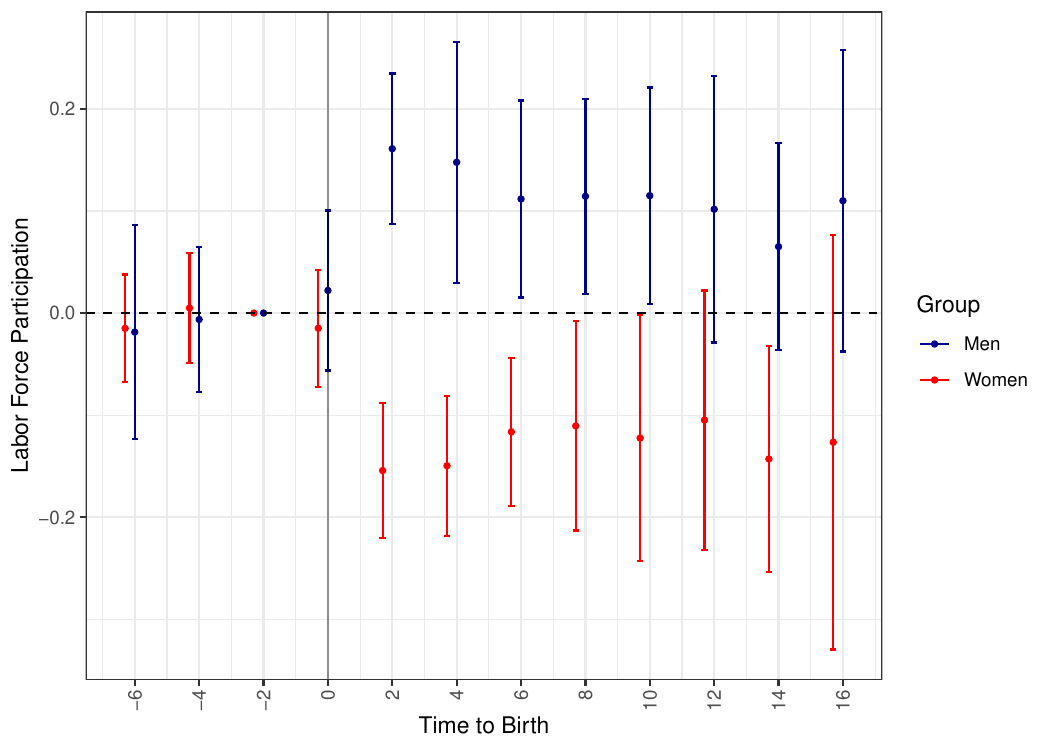} 
        \caption{Law States Only} 
    \end{subfigure}%
    \hfill
    \begin{subfigure}{.48\textwidth}
        \centering
        \includegraphics[width=\linewidth]{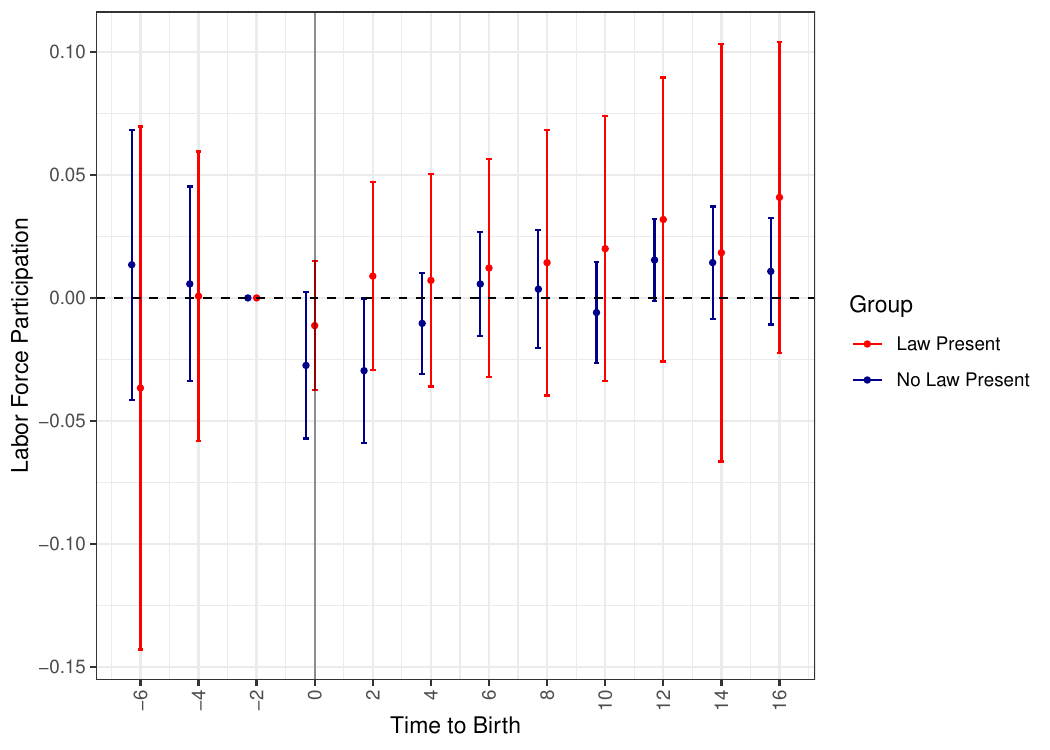} 
        \caption{Women Only} 
    \end{subfigure}
    \label{fig:birtheventLFP}
{\footnotesize \justifying \singlespacing{This figure depicts results from an interacted regression around the timing of the first child birth, with 95\% confidence intervals. \textbf{Panel A} shows the comparison of LFP by gender following the birth of the first child for families living in states with breastfeeding legislation, and  \textbf{Panel B} the comparison between women who live in states with legislation and women who do not live in states with legislation. Data is taken from the PSID and childbirth event calculated by \cite{cortes2023children}. $\ell = -2$ is normalized to zero to account for multicollinearity between relative and calendar time.} \par}

\end{figure}

Results for this analysis are depicted in Figure \ref{fig:birtheventLFP}. First I study the difference between the probability of labor force participation in men and women who all reside in law enacting states, depicted in Panel A. In this setting, the relative timing indicator for before and after birth is interacted with a binary gender variable. Consistent with \cite{cortes2023children} who find a significant decline in wage earnings post-birth for mothers, I find a significant drop in LFP of women compared to men, even in states where the law is in effect. The probability of LFP by men increases after birth, in contrast. These differences dissipate by the child's later years of life, becoming non-significant by age 10. In Panel B, I plot results for a specification where the interaction term is an indicator for living in a breastfeeding law state at the time of birth, and the sample itself is restricted entirely to women. That is, this panel compares the different effects of the legislation on women who live in legislation-enacting states to women who live elsewhere. These results show that for those women who do not live in states with the breastfeeding law, the significant decrease in probability of labor force participation is exhibited in the first 2 years after birth. In contrast, women who live in states where the breastfeeding law is in place do not exhibit a significant reduction in the probability of LFP following the birth. Across the later years of the child's life, the point estimates for women in the law-enacting states are consistently larger than the estimates for those in the non-law-enacting states (although they are not significant). These results contribute to the overall findings of this paper by weakly suggesting that women in the law-enacting states are less likely to drop out of the workforce following the childbirth event.

\subsection{Heterogeneity Analysis}\label{subsection:het_analysis}

In this section, I examine the effects of the legislation on FLFP by individual heterogeneous characteristics. Specifically, I examine occupation types, race, and education level. 

\subsubsection{By Occupation}

\begin{figure}[H]
\caption{Results by Occupational Category}
    \centering
    \begin{subfigure}{0.3\textwidth}
        \centering
        \includegraphics[width=\linewidth]{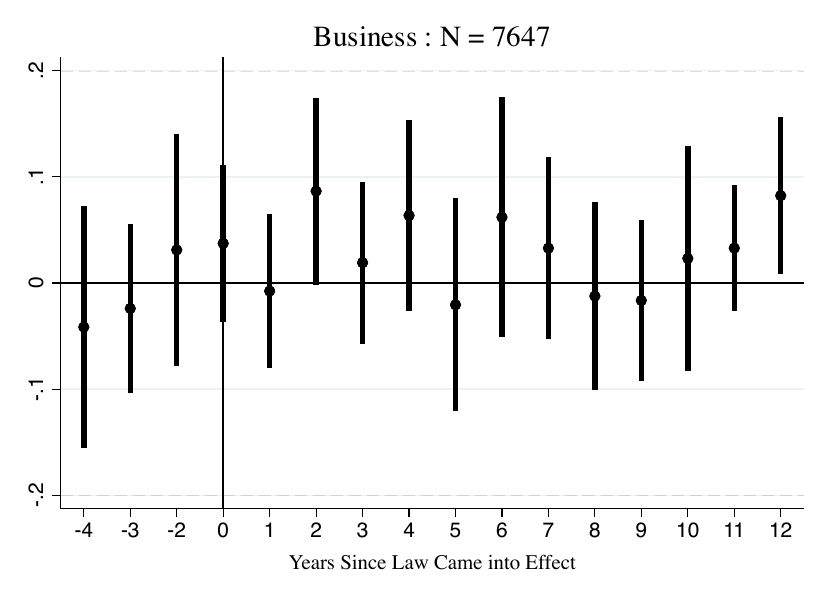}
        \caption{Business}
    \end{subfigure}
    \hfill
    \begin{subfigure}{0.3\textwidth}
        \centering
        \includegraphics[width=\linewidth]{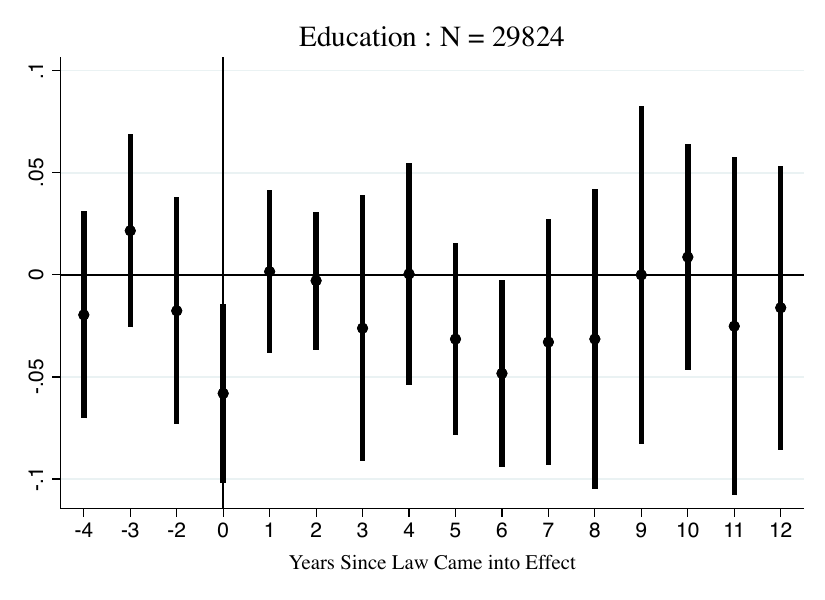}
        \caption{Education}
    \end{subfigure}
    \hfill
    \begin{subfigure}{0.3\textwidth}
        \centering
        \includegraphics[width=\linewidth]{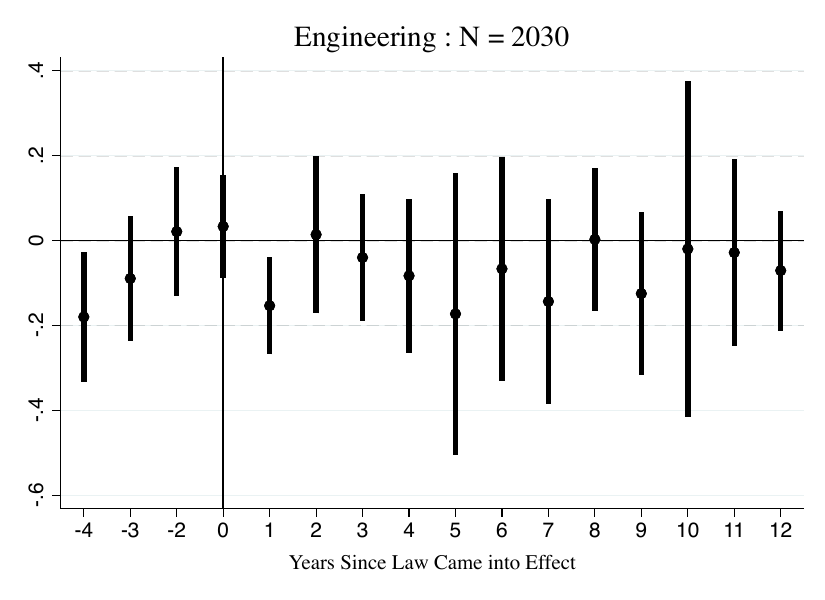}
        \caption{Engineering}
    \end{subfigure}

    \vspace{1em} 
    \begin{subfigure}{0.3\textwidth}
        \centering
        \includegraphics[width=\linewidth]{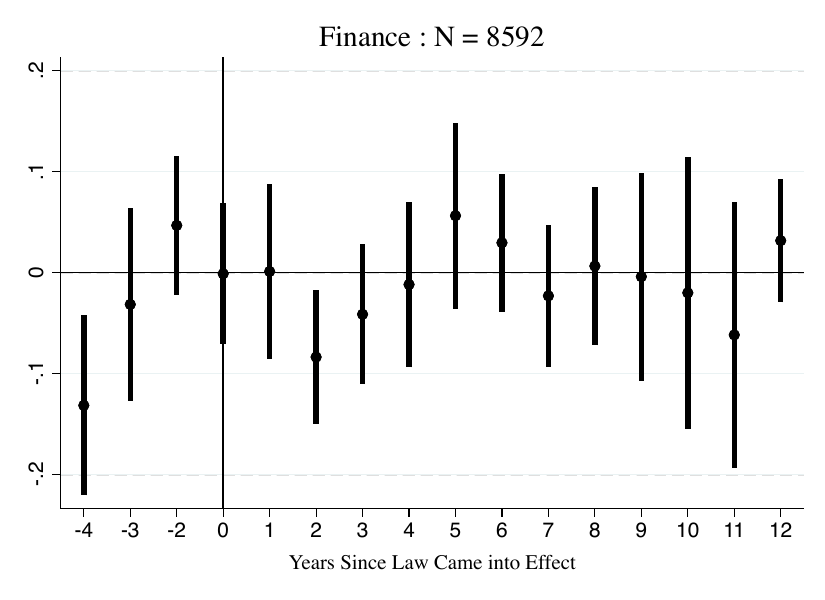}
        \caption{Finance}
    \end{subfigure}
    \hfill
    \begin{subfigure}{0.3\textwidth}
        \centering
        \includegraphics[width=\linewidth]{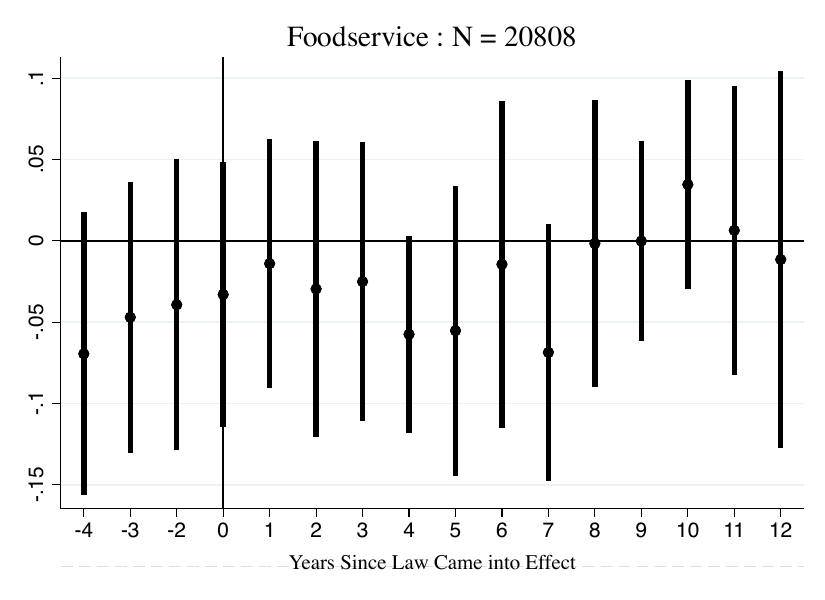}
        \caption{Food Service}
    \end{subfigure}
    \hfill
    \begin{subfigure}{0.3\textwidth}
        \centering
        \includegraphics[width=\linewidth]{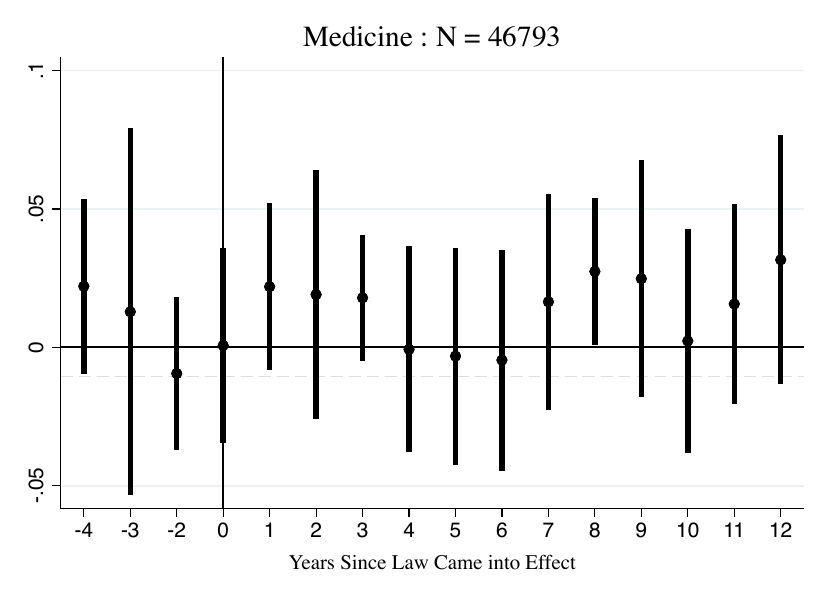}
        \caption{Medicine}
    \end{subfigure}

    \vspace{1em}
    \begin{subfigure}{0.3\textwidth}
        \centering
        \includegraphics[width=\linewidth]{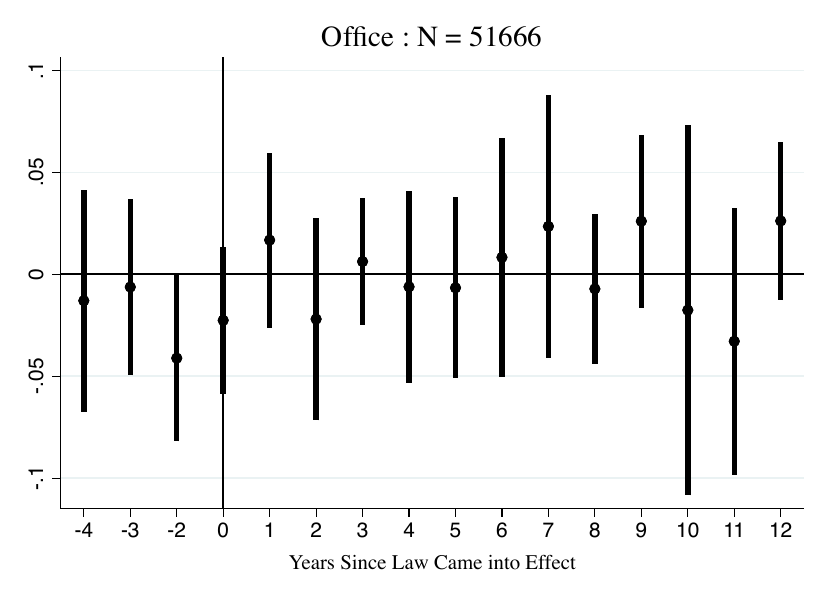}
        \caption{Office Work (General)}
    \end{subfigure}
    \hfill
    \begin{subfigure}{0.3\textwidth}
        \centering
        \includegraphics[width=\linewidth]{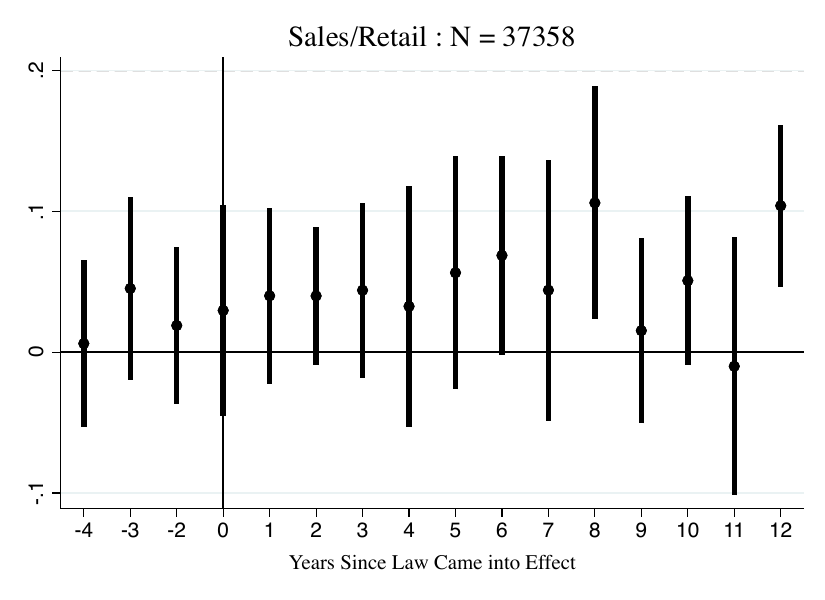}
        \caption{Sales/Retail}
    \end{subfigure}
    \hfill
    \begin{subfigure}{0.3\textwidth}
        \centering
        \includegraphics[width=\linewidth]{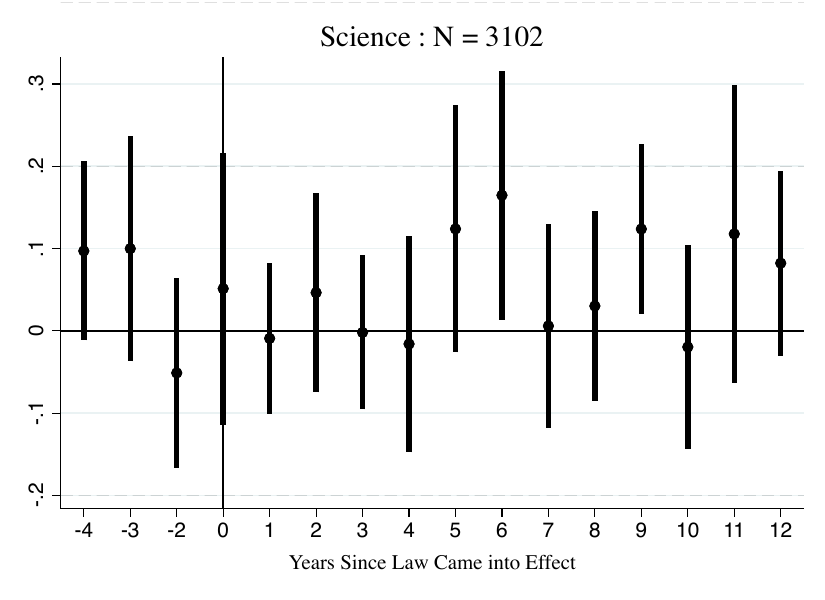}
        \caption{Science}
    \end{subfigure}

    \vspace{1em}
    \begin{subfigure}{0.3\textwidth}
        \centering
        \includegraphics[width=\linewidth]{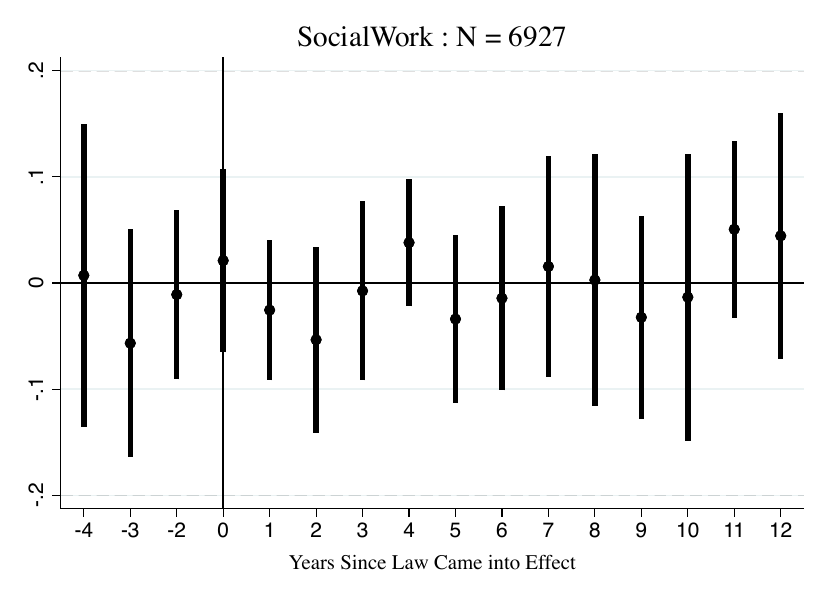}
        \caption{Social Work}
    \end{subfigure}
    \hfill
    \begin{subfigure}{0.3\textwidth}
        \centering
        \includegraphics[width=\linewidth]{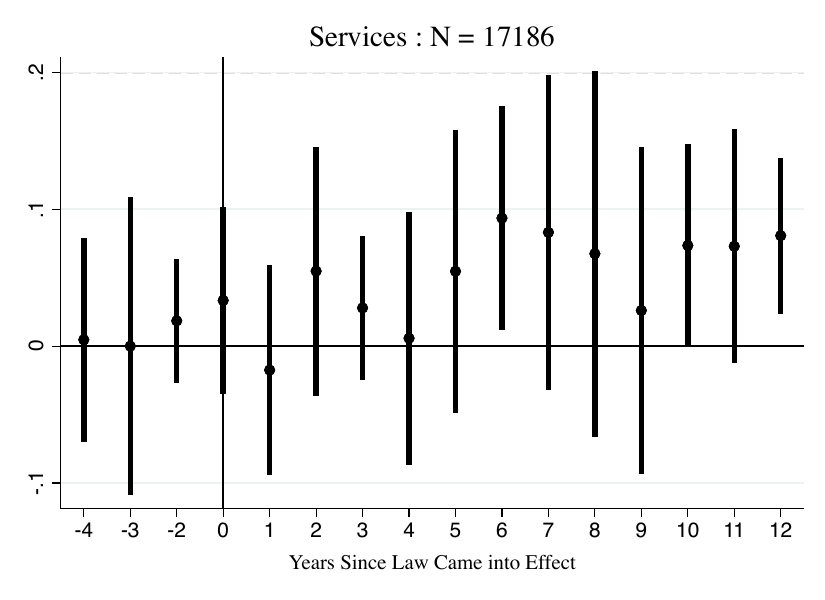}
        \caption{Services (non-food)}
    \end{subfigure}
    \hfill
    \begin{subfigure}{0.3\textwidth}
        \centering
        \includegraphics[width=\linewidth]{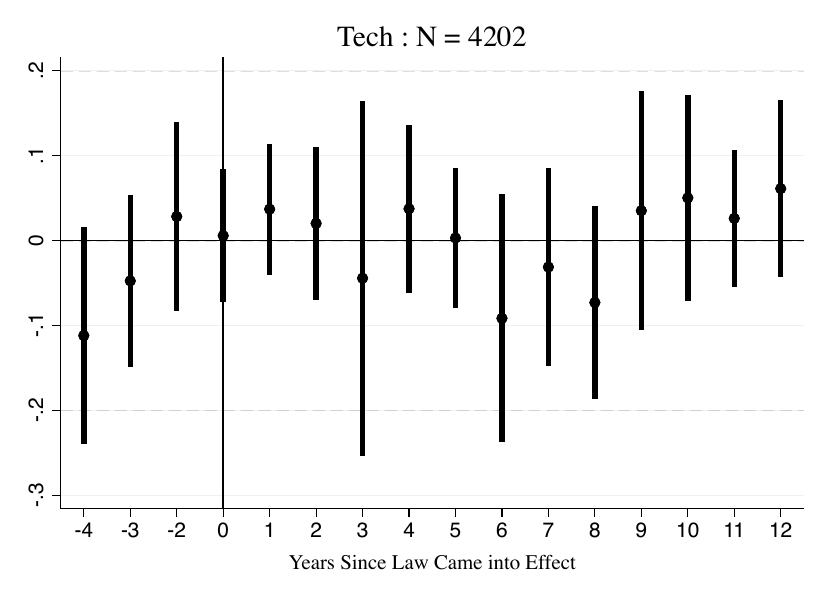}
        \caption{Tech}
    \end{subfigure}
    \label{fig:OCCP_subfigures}
    {\footnotesize \justifying \singlespacing{This figure depicts results 2WFE regressions with 95\% confidence intervals. Data is taken from the ACS sample of women who gave birth in the previous calendar year. Each panel depicts the results for the data subsetted by the indicated occupation type. Time $\ell = -1$ is normalized to zero to account for multicollinearity between relative and calendar time. Standard errors are clustered at the state level.  } \par}
\end{figure}

The ACS data contains a large number of observations with rich information about the occupation and industry of each respondent. Restricting the sample to subsets of individuals in each major occupation code still allows for the analysis to utilize a large number of observations. I plot the event study estimates and their 95\% confidence intervals in Figure \ref{fig:OCCP_subfigures}, for the twelve largest occupational groups in the ACS data. 

Results indicate substantial effect heterogeneity by occupational category. The steady increase in the probability of FLFP evidenced in the dynamic ATT results for the entire sample are reflected in the subsets of women who work in science, sales/retail, and non-food services. There are additional, but less persistent, positive impacts of the legislation on women working in medicine. The strongest \textit{negatively} impacted occupations include education, engineering, and finance, the latter two being highly male-dominated industries which would be consistent with a negative spillover effect. Education is traditionally a female-dominated occupation, however, the scheduling demands of this occupation may be incompatible with taking breaks to breastfeed or express milk. 

\subsubsection{By Education Level}

\begin{figure}[H]
    \centering
    \caption{College Education}
    \begin{subfigure}{.48\textwidth}
        \centering
        \includegraphics[width=\linewidth]{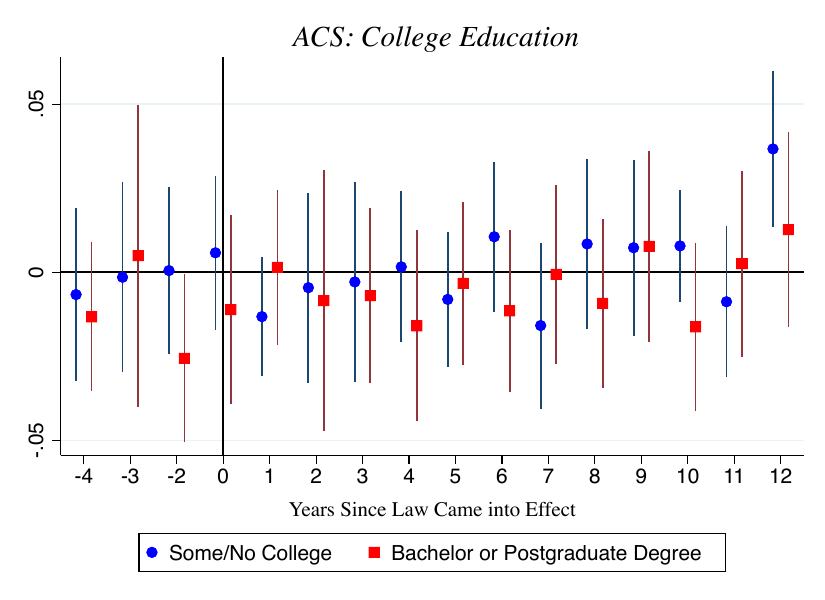} 
        \caption{ACS} 
    \end{subfigure}%
    \hfill
    \begin{subfigure}{.48\textwidth}
        \centering
        \includegraphics[width=\linewidth]{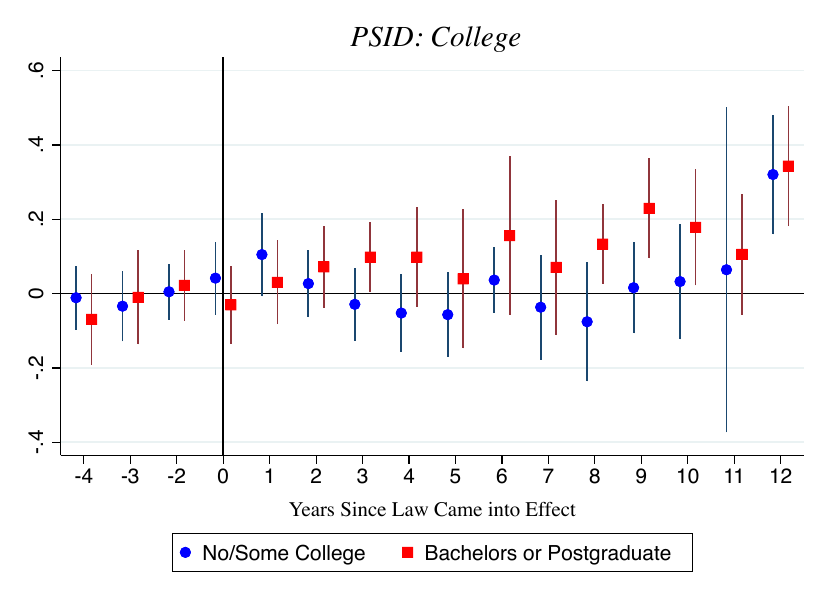} 
        \caption{PSID} 
    \end{subfigure}
    \label{fig:college}
{\footnotesize \justifying \singlespacing{This figure depicts results 2WFE regressions with 95\% confidence intervals. \textbf{Panel A} shows results using the ACS 2005-2019 sample of women who gave birth in the previous year, and \textbf{Panel B} the PSID from 1997-2021 of women with children under age 3. Time $\ell = -1$ is normalized to zero to account for multicollinearity between relative and calendar time. Standard errors are clustered at the state level.  } \par}
\end{figure}

As educational attainment is known to impact both the decision to work and to breastfeed (see Section \ref{section:background}), I perform a heterogeneity analysis by higher education status in this section. Results are depicted in Figure \ref{fig:college}.

There are major differences in the results between the two datasets in this analysis. The stronger impacts of the legislation on the probability of FLFP in Panel B, which uses the PSID data, is consistent with the findings in \cite{del2012does}, who only find impacts of UK breastfeeding workplace amenities on FLFP for women with a college-level education. However, in the ACS data, there are no significant differences by education level. The two datasets differ noticeably in educational characteristics, see Table \ref{tab:sumstats_ACS_PSID}. 

\subsubsection{By Race/Ethnicity}

\begin{figure}[H]
    \centering
    \caption{White vs Black Women}
    \begin{subfigure}{.48\textwidth}
        \centering
        \includegraphics[width=\linewidth]{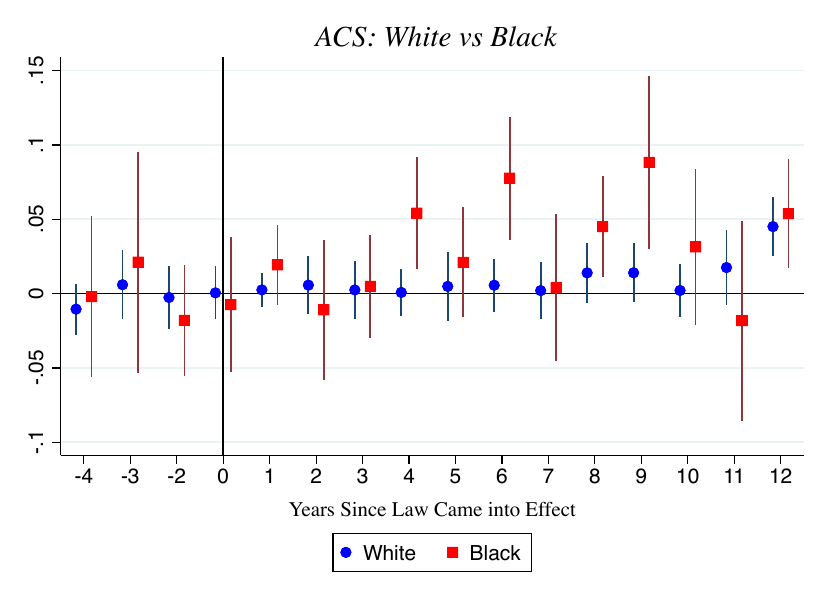} 
        \caption{ACS} 
    \end{subfigure}%
    \hfill
    \begin{subfigure}{.48\textwidth}
        \centering
        \includegraphics[width=\linewidth]{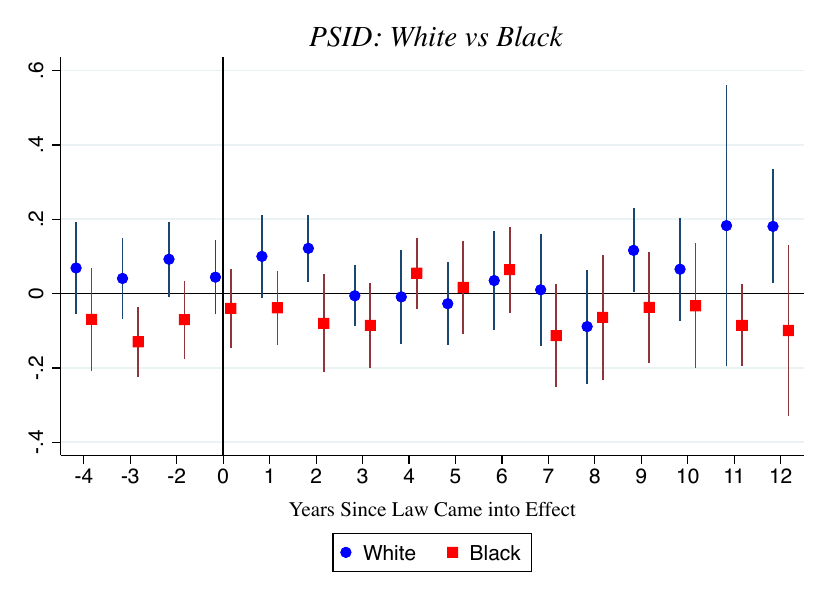} 
        \caption{PSID} 
    \end{subfigure}
    \label{fig:whiteblack}
{\footnotesize \justifying \singlespacing{This figure depicts results 2WFE regressions with 95\% confidence intervals. \textbf{Panel A} shows results using the ACS 2005-2019 sample of women who gave birth in the previous year, and \textbf{Panel B} the PSID from 1997-2021 of women with children under age 3. Time $\ell = -1$ is normalized to zero to account for multicollinearity between relative and calendar time. Point estimates are shown in table \ref{tab:hetbyrace}. Standard errors are clustered at the state level.  } \par}

\end{figure}

Finally, I subset the data to examine effects of the breastfeeding legislation by race and ethnicity. First, I compare White and Black women, in both datasets. The ACS and PSID exhibit very different compositions of racial demographics (see Table \ref{tab:sumstats_ACS_PSID}), which is reflected in the results depicted in Figure \ref{fig:whiteblack}.

According to the ACS data (Panel A), there is evidence of a strong impact on the likelihood of FLFP for Black women in the sample, beginning as early as 4 years after implementation of the law. The magnitude of these coefficients is quite high, above 5 percentage points. However, the PSID sample (Panel B) does not exhibit the same trend; there, it is White women who appear to be more strongly impacted by the law. Although this evidence may be inconclusive, it points to some significant heterogeneity in the effects of the law by race. 

\begin{figure}[H]
\centering
\begin{minipage}{0.65\textwidth} 
\caption{Hispanic/Latina Women}
\label{fig:hispanic}
\includegraphics[width=10cm]{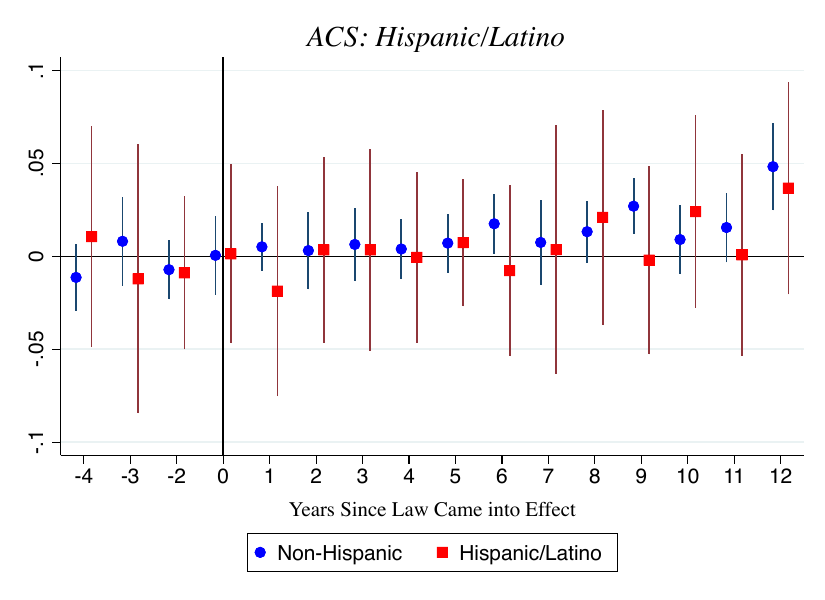}
{\footnotesize \justifying \singlespacing{This figure depicts results for 2WFE regressions with data subsetted by hispanic/latino ethnic status, with 95\% confidence intervals using the ACS 2005-2019 main sample, which consists of women who gave birth in the last year. Standard errors are clustered at the state level. } \par}
\end{minipage}
\end{figure}

Finally, I wish to examine whether there are differential effects based on Hispanic or Latino origin. The PSID does not allow for clean identification of this group of women, so I only show results using the ACS in Figure \ref{fig:hispanic}. The slowly increasing trend of the impact of the breastfeeding legislation is only evidenced here for non-Hispanic/Latina women. This is in line with prior literature showing that Latina women are much less likely to breastfeed, and therefore would be less impacted by the law \citep{wouk2016clinical}. 

\subsection{Robustness Tests}\label{subsection:robtests}

This section includes a suite of robustness tests and other extensions to examine the validity of the earlier findings.

\subsubsection{Influence of historic FLFP rates}

One major identification concern in this paper is whether states may be passing their breastfeeding legislation in response to low or declining maternal labor force participation rates. If declining LFP rates prompted lawmakers to pass such laws in response, any subsequent rise in LFP could be mistaken for a causal effect of the laws when it might just be a result of regression to the mean. One reassuring aspect of the main results is the lack of significant pre-trends. However, to further address this concern, I repeat the analysis and subset the data by highest and lowest terciles of \textit{pre-law} labor force participation rates. That is, I calculate the average FLFP rate by state in the years prior to the law being passed, and then subset the data by terciles of this average rate, performing analysis separately for the highest and lowest tercile. Results are depicted in Figure \ref{fig:highvlowLFP}, with point estimates in Appendix Table \ref{tab:highvlow_LFP}.

\begin{figure}[H]
    \centering
    \caption{High vs. Low LFP States}
    \begin{subfigure}{.48\textwidth}
        \centering
        \includegraphics[width=\linewidth]{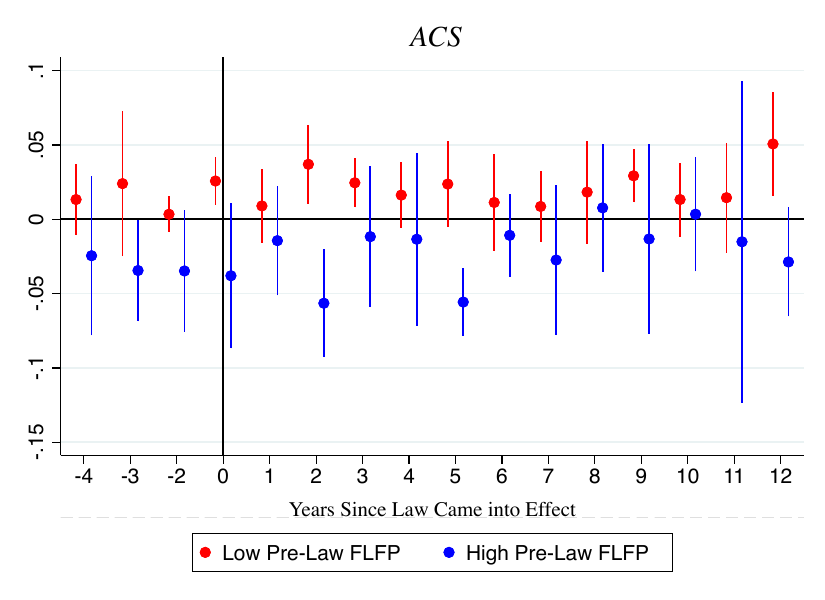} 
        \caption{ACS: recent birth} 
    \end{subfigure}%
    \hfill
    \begin{subfigure}{.48\textwidth}
        \centering
        \includegraphics[width=\linewidth]{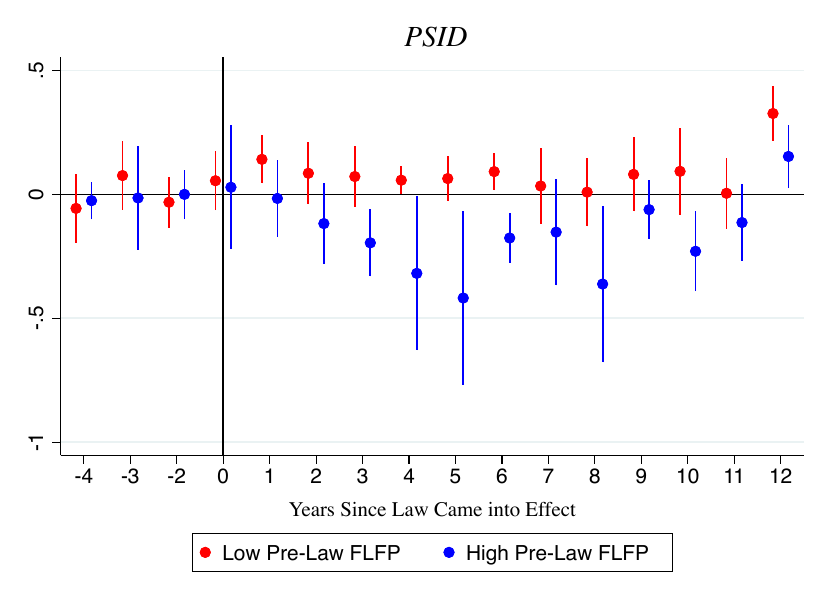} 
        \caption{PSID: child $< 3$ } 
    \end{subfigure}
    \label{fig:highvlowLFP}
{\footnotesize \justifying \singlespacing{This figure depicts results from the main regression with 95\% confidence intervals. \textbf{Panel A} shows the comparison of LFP between higher and lower terciles of pre-law LFP using ACS data, and  \textbf{Panel B} with PSID data. Standard errors are clustered at the state level. $\ell = -1$ is normalized to zero to account for multicollinearity between relative and calendar time.} \par}

\end{figure}

Across both datasets, there is a clear difference in the impact of the law for those states in the lowest (red) versus highest (blue). Reassuringly, there is still a lack of pre-trends for the estimates even when examining those in the lowest tercile of pre-law LFP. However, results indicate a highly heterogeneous effect of the laws based on historic levels of female LFP, suggesting that the impact of the laws depend on pre-existing labor market conditions. Since we observe a positive impact of the law in low FLFP states, this suggests that the law had a potential causal effect on increasing female labor force participation in areas where participation was historically low. The significant positive effects for the low FLFP states occur much earlier in the ACS data compared to the main results (at 0-2 years compared to 6-7 years in Figure \ref{fig:crosssectionLFP}). In high FLFP states, where participation was already high, there is evidence of a negative impact of the laws. This negative effect might suggest that the law did not address the specific constraints faced by these women and in fact, it may have backfired in these states. Taken together, these findings imply that the law’s effectiveness depended on the initial state of the labor market, while the lack of pre-trends further supports the assumption that timing of these laws are likely to be exogenous to historic FLFP rates.

\subsubsection{Placebo Tests}

To check the validity of the estimates for the cross-sectional data using women who have recently given birth (ACS data), or who have children present under age 3 (PSID), I perform a panel regression using the PSID data for various samples of women who, ex-ante, should not be affected by the breastfeeding laws. I repeat the analysis for women who are over child-bearing age of 45, women who only have children older than age 6 in the household, women who have no children in the household, and women who \textit{never} have children across the entire sample period. I use the PSID data as I am able to compare results for both state and person-level fixed effects, as the various subsets of the main survey data do not restrict the panel structure in terms of years present in the sample. 

Results are depicted in Appendix Table \ref{tab:placebos}. For the most part, there is no impact of the breastfeeding law on these groups of women. Older women exhibit no impact of the law, as expected. However, in specifications with the individual-level fixed effects, the puzzling finding from the interacted panel regressions re-emerges, where there appears to be a \textit{negative} effect of the legislation on women who have older children, or who never have children across the sample period. This finding might not necessarily discredit the main result of this paper, which finds a positive impact of the legislation on the FLFP of mothers with young children, but rather it could be uncovering a spillover effect of the law on other groups of women. Some employers could feel burdened by the requirement to provide breastfeeding accommodations and pass on costs to all employees (e.g., by reducing work flexibility for others, lowering wages, or increasing workload for childless employees). More concerningly, employers might indirectly treat all women of childbearing age differently following the introduction of breastfeeding laws, out of concern that they may eventually need to offer accommodations - even if an employee is currently childless. These findings raise important questions for future research. 

\subsubsection{Influence of Related Policies}

To address concerns about the possible confounding effects of other laws regarding maternity and the workplace, I repeat the analysis including a binary indicator for whether or not a paid leave provision or job protection law is currently in place in each state at each time $t$. I report results in Appendix Table \ref{tab:otherpolicies}. Results using the ACS data are robust to inclusion of either or both of these controls. The PSID findings weaken slightly when paid leave laws are included as a control, but not when the job protection laws are added as a control. These differences could be due to significant effect heterogeneity in terms of college education, whereby the breastfeeding laws seemed more impactful for women with college education in the PSID sample, as well as compositional differences across race and education levels compared to the ACS data (see Table \ref{tab:sumstats_ACS_PSID}). Further research is needed to ideally study the combined effect of these laws. Taken together, the overall findings are mostly robust to controlling for alternative maternity related workplace-policies.

\subsubsection{Other Robustness Tests}

In addition to checking robustness of the main results to a variety of specifications of the standard errors\footnote{taking into account the fact that a small number of clusters may over reject the null as in \cite{cameron2008bootstrap}} (see Appendix Tables \ref{tab:ACS_results} and \ref{tab:PSID_results}), I perform a leave-one-out analysis to ensure that the results are not driven by the influence of any particular individual state. Results are depicted in the Appendix and show that the ACS data results are robust to removing any single state, with slight changes in magnitudes of the coefficients. The only slight exceptions are in the cases when New York or Colorado are removed from the analysis, where the significant effect in the 6th year after the law is passed becomes reduced - however, the effect becomes significant again in year 8, which is consistent with the main results. Finally, I check whether the passing of the ``weak" laws, which only require workplace provisions for public employees, has an impact on the FLFP of public/state employees. I do this by removing all women who do not indicate themselves as public employees in the ACS data, and repeating the analysis, and removing the other law states, so that the treatment group is the public employee law state and the control group is the states without any legislation. The results are depicted in the Appendix (Figure A1). Findings are mixed, with both strongly positive and negative impacts of the laws in the subsequent years after implementation, with no discernable pattern. This could simply be due to the small number of treated groups (two).

\section{Conclusion}

This paper demonstrates evidence of a positive impact of workplace breastfeeding legislation on female labor force participation for mothers of young children, across a variety of datsets and empirical models. Analysis using ACS data shows a slow and steady increase in impact of the laws, beginning around 1.5 percentage points increase in the liklihood of FLFP 6 years after law implementation and increasing to 2.4 percent by year 9. Using PSID data in the same econometric framework shows a more immediate impact of the legislation, increasing the likelihood of FLFP by 5.9 percentage points in the second year after implementation, but the effect does not seem to persist. I find evidence of significant effect heterogeneity by race, educational status, and occupation, but the specifics of these findings are highly dependent upon which dataset is being used for the analysis. I also find that the laws were much more impactful for those states who were in the bottom tercile of FLFP prior to the law being enacted (with no evidence of pre-trends). Finally, I uncover a puzzling finding whereby women without children, and women with older children, seem to be persistently and significantly negatively impacted by the law, reducing their likelihood of LFP by 3 - 4 percentage points following passing of the laws.

If optimal amount of breastfeeding is chosen so that marginal benefits equal the marginal costs (in terms of labor income, for example), perhaps women who return to work after implementation of the laws experience realized personal costs of workplace pumping and breastfeeding that are greater than expected. If the experienced benefits of workplace breastfeeding are not as significant as most women are led to believe, or if it inherently difficult to pump or breastfeed at work regardless of the provision of workplace breastfeeding amenities, it is understandable that the effects of workplace breastfeeding legislation would not be persistent ex-post, as appears to be the case in the PSID sample. Indeed, \cite{chen2006effects} find that despite ample provision of workplace breastfeeding amenities, Taiwanese mothers are not very likely to continue to breastfeed after returning to work. \cite{felice2017breastfeeding} show that working mothers report feeling ``isolated and embarassed" for engaging in workplace pumping. Equally concerning are the possible negative spillover effects for women without young children, who may be anticipated to make use of the breastfeeding amenities or assigned higher workloads to make up for the workloads of those women who do utilise them. 

It should be noted, however, that women in the IFPS II sample are much more likely to report high job satisfaction if they reside in a state with the workplace breastfeeding policies in place. This demonstrates a potential positive spillover effect - even though working mothers may not make significantly more use of the workplace breastfeeding amenities, their very provision can change the worker's perception of her firm, which may increase her likelihood to return to her job after giving birth. 

As shown with the large ACS data, we would expect that any benefits to returning to work after birth with these amenities in place would in fact be persistent. The transience evidenced in the PSID (except for college educated women) could simply be influenced by the oversampling of poorer, minority-headed households in that survey. The biennial aspect of the survey may also be leading to imprecision in the assignment of treatment time. Determining whether the supply or demand side of the labor market is truly responsible for a transient or persistent effect is beyond the scope of this paper without having detailed information about the rationale behind participation decisions of women. Survey evidence, however, indicates that work schedule flexibility is related to workplace breastfeeding continuation \citep{felice2017breastfeeding}, which suggests that the presence of breastfeeding legislation alone, without other family-friendly policies in place, may not be enough to support the full-time labor force participation of mothers. Robustness checks in the paper showing a possible relationship between paid maternity leave laws and the breastfeeding law impacts supports this idea. Other research should examine the impacts of recent hospital shifts towards Baby-Friendly practices on female labor force participation \citep{nelson2019trends}. 

Future work may consider undertaking similar analysis using aggregate measures of women's LFP as outcomes of interest, to connect this study to a larger macroeconomic perspective. It is also important to investigate whether the negative effect on childless women, and women with older children, is more pronounced in certain industries or workplaces that might be more impacted by breastfeeding accommodations. For instance, sectors with more rigid work schedules or limited flexibility might see a larger spillover effect on childless women. I leave this avenue of exploration to future work. As family-friendly workplace policies may carry an implicit wage cost \citep{heywood2007implicit}, examination of wage effects is another important extension. The findings regarding additional impacts of paid maternal leave laws in robustness checks warrants further investigation, as the combined treatments of both paid leave and breastfeeding accomodation may have differential effects - however, as there are very few states with robust paid maternity leave laws, this remains a difficult avenue of study. One exciting possibility to further explore evidence of heterogeneous treatment effects uncovered in this paper could be to utilize recently proposed machine-learning based methods to study drivers of treatment effect heterogeneity in the staggered treatment adoption setting \citep{gavrilova2023dynamic, hatamyar2023machine}. These, and other robust nonparametric methods, could be used to investigate the puzzling finding of possible negative spillover effects of the legislation on the labor force participation of childless women. They could also be used to ensure robustness of the results indicating a stronger effect of the laws in states with lower rates of FLFP prior to implementation. 

\bibliographystyle{apalike}
\bibliography{bf}

\begin{thebibliography}{}

\bibitem[Abdulwadud and Snow, 2007]{abdulwadud2007interventions}
Abdulwadud, O.~A. and Snow, M.~E. (2007).
\newblock Interventions in the workplace to support breastfeeding for women in employment.
\newblock {\em Cochrane database of systematic reviews}, (3).

\bibitem[Adda et~al., 2017]{adda2017career}
Adda, J., Dustmann, C., and Stevens, K. (2017).
\newblock The career costs of children.
\newblock {\em Journal of Political Economy}, 125(2):293--337.

\bibitem[Athey and Imbens, 2018]{athey2018design}
Athey, S. and Imbens, G.~W. (2018).
\newblock Design-based analysis in difference-in-differences settings with staggered adoption.
\newblock Technical report, National Bureau of Economic Research.

\bibitem[Baker and Milligan, 2008]{baker2008does}
Baker, M. and Milligan, K. (2008).
\newblock How does job-protected maternity leave affect mothers’ employment?
\newblock {\em Journal of Labor Economics}, 26(4):655--691.

\bibitem[Becker and Becker, 2009]{becker2009treatise}
Becker, G.~S. and Becker, G.~S. (2009).
\newblock {\em A Treatise on the Family}.
\newblock Harvard university press.

\bibitem[Bertrand et~al., 2010]{bertrand2010dynamics}
Bertrand, M., Goldin, C., and Katz, L.~F. (2010).
\newblock Dynamics of the gender gap for young professionals in the financial and corporate sectors.
\newblock {\em American economic journal: applied economics}, 2(3):228--55.

\bibitem[Blau and Kahn, 2013]{blau2013female}
Blau, F.~D. and Kahn, L.~M. (2013).
\newblock Female labor supply: Why is the united states falling behind?
\newblock {\em American Economic Review}, 103(3):251--56.

\bibitem[Blau and Kahn, 2017]{blau2017gender}
Blau, F.~D. and Kahn, L.~M. (2017).
\newblock The gender wage gap: Extent, trends, and explanations.
\newblock {\em Journal of economic literature}, 55(3):789--865.

\bibitem[Borusyak and Jaravel, 2017]{borusyak2017revisiting}
Borusyak, K. and Jaravel, X. (2017).
\newblock Revisiting event study designs.
\newblock {\em Available at SSRN 2826228}.

\bibitem[Callaway et~al., 2018]{callaway2018difference}
Callaway, B., Sant’Anna, P.~H., et~al. (2018).
\newblock Difference-in-differences with multiple time periods and an application on the minimum wage and employment.
\newblock {\em arXiv preprint arXiv:1803.09015}, pages 1--47.

\bibitem[Cameron et~al., 2008]{cameron2008bootstrap}
Cameron, A.~C., Gelbach, J.~B., and Miller, D.~L. (2008).
\newblock Bootstrap-based improvements for inference with clustered errors.
\newblock {\em The review of economics and statistics}, 90(3):414--427.

\bibitem[Chatterji and Frick, 2005]{chatterji2005does}
Chatterji, P. and Frick, K.~D. (2005).
\newblock Does returning to work after childbirth affect breastfeeding practices?
\newblock {\em Review of Economics of the Household}, 3(3):315--335.

\bibitem[Chen et~al., 2006]{chen2006effects}
Chen, Y.~C., Wu, Y.-C., and Chie, W.-C. (2006).
\newblock Effects of work-related factors on the breastfeeding behavior of working mothers in a taiwanese semiconductor manufacturer: a cross-sectional survey.
\newblock {\em BMC Public Health}, 6(1):160.

\bibitem[Cort{\'e}s and Pan, 2023]{cortes2023children}
Cort{\'e}s, P. and Pan, J. (2023).
\newblock Children and the remaining gender gaps in the labor market.
\newblock {\em Journal of Economic Literature}, 61(4):1359--1409.

\bibitem[De~Chaisemartin and d'Haultfoeuille, 2020]{de2020two}
De~Chaisemartin, C. and d'Haultfoeuille, X. (2020).
\newblock Two-way fixed effects estimators with heterogeneous treatment effects.
\newblock {\em American Economic Review}, 110(9):2964--96.

\bibitem[Del~Bono and Pronzato, 2012]{del2012does}
Del~Bono, E. and Pronzato, C.~D. (2012).
\newblock Does breastfeeding support at work help mothers and employers at the same time?
\newblock Technical report, ISER Working Paper Series.

\bibitem[Dubois and Girard, 2003]{dubois2003social}
Dubois, L. and Girard, M. (2003).
\newblock Social determinants of initiation, duration and exclusivity of breastfeeding at the population level.
\newblock {\em Canadian journal of public health}, 94(4):300--305.

\bibitem[Fallon et~al., 2019]{fallon2019impact}
Fallon, V.~M., Harrold, J.~A., and Chisholm, A. (2019).
\newblock The impact of the uk baby friendly initiative on maternal and infant health outcomes: A mixed-methods systematic review.
\newblock {\em Maternal \& Child Nutrition}, 15(3):e12778.

\bibitem[Fein and Roe, 1998]{fein1998effect}
Fein, S.~B. and Roe, B. (1998).
\newblock The effect of work status on initiation and duration of breast-feeding.
\newblock {\em American journal of public health}, 88(7):1042--1046.

\bibitem[Felice et~al., 2017]{felice2017breastfeeding}
Felice, J.~P., Geraghty, S.~R., Quaglieri, C.~W., Yamada, R., Wong, A.~J., and Rasmussen, K.~M. (2017).
\newblock “breastfeeding” without baby: A longitudinal, qualitative investigation of how mothers perceive, feel about, and practice human milk expression.
\newblock {\em Maternal \& child nutrition}, 13(3):e12426.

\bibitem[Gault et~al., 2014]{gault2014paid}
Gault, B., Hartmann, H., Hegewisch, A., Milli, J., and Reichlin, L. (2014).
\newblock Paid parental leave in the united states: What the data tell us about access, usage, and economic and health benefits.

\bibitem[Gavrilova et~al., 2023]{gavrilova2023dynamic}
Gavrilova, E., Lang{\o}rgen, A., and Zoutman, F.~T. (2023).
\newblock Dynamic causal forests, with an application to payroll tax incidence in norway.
\newblock Technical report, CESifo Working Paper.

\bibitem[Goodman-Bacon, 2018]{goodman2018difference}
Goodman-Bacon, A. (2018).
\newblock Difference-in-differences with variation in treatment timing.
\newblock Technical report, National Bureau of Economic Research.

\bibitem[Guendelman et~al., 2009]{guendelman2009juggling}
Guendelman, S., Kosa, J.~L., Pearl, M., Graham, S., Goodman, J., and Kharrazi, M. (2009).
\newblock Juggling work and breastfeeding: effects of maternity leave and occupational characteristics.
\newblock {\em Pediatrics}, 123(1):e38--e46.

\bibitem[Hatamyar et~al., 2023]{hatamyar2023machine}
Hatamyar, J., Kreif, N., Rocha, R., and Huber, M. (2023).
\newblock Machine learning for staggered difference-in-differences and dynamic treatment effect heterogeneity.
\newblock {\em arXiv preprint arXiv:2310.11962}.

\bibitem[Hatsor and Shurtz, 2019]{hatsor2019breastfeeding}
Hatsor, L. and Shurtz, I. (2019).
\newblock Breastfeeding and labor supply of new mothers: Evidence from a baby formula hazard realization.

\bibitem[Hauck et~al., 2020]{hauck2020integrating}
Hauck, K., Miraldo, M., and Singh, S. (2020).
\newblock Integrating motherhood and employment: A 22-year analysis investigating impacts of us workplace breastfeeding policy.
\newblock {\em SSM-Population Health}, page 100580.

\bibitem[Heywood et~al., 2007]{heywood2007implicit}
Heywood, J.~S., Siebert, W.~S., and Wei, X. (2007).
\newblock The implicit wage costs of family friendly work practices.
\newblock {\em Oxford Economic Papers}, 59(2):275--300.

\bibitem[Imai and Kim, 2020]{imai2020use}
Imai, K. and Kim, I.~S. (2020).
\newblock On the use of two-way fixed effects regression models for causal inference with panel data.
\newblock {\em Political Analysis}, pages 1--11.

\bibitem[Kleven et~al., 2019]{kleven2019children}
Kleven, H., Landais, C., and S{\o}gaard, J.~E. (2019).
\newblock Children and gender inequality: Evidence from denmark.
\newblock {\em American Economic Journal: Applied Economics}, 11(4):181--209.

\bibitem[Kurinij et~al., 1989]{kurinij1989does}
Kurinij, N., Shiono, P.~H., Ezrine, S.~F., and Rhoads, G.~G. (1989).
\newblock Does maternal employment affect breast-feeding?
\newblock {\em American Journal of Public Health}, 79(9):1247--1250.

\bibitem[Kuziemko et~al., 2018]{kuziemko2018mommy}
Kuziemko, I., Pan, J., Shen, J., and Washington, E. (2018).
\newblock The mommy effect: Do women anticipate the employment effects of motherhood?
\newblock Technical report, National Bureau of Economic Research.

\bibitem[Lubold, 2016]{lubold2016breastfeeding}
Lubold, A.~M. (2016).
\newblock Breastfeeding and employment: A propensity score matching approach.
\newblock {\em Sociological Spectrum}, 36(6):391--405.

\bibitem[Mandal et~al., 2014]{mandal2014work}
Mandal, B., Roe, B.~E., and Fein, S.~B. (2014).
\newblock Work and breastfeeding decisions are jointly determined for higher socioeconomic status us mothers.
\newblock {\em Review of Economics of the Household}, 12(2):237--257.

\bibitem[Murtagh and Moulton, 2011]{murtagh2011working}
Murtagh, L. and Moulton, A.~D. (2011).
\newblock Working mothers, breastfeeding, and the law.
\newblock {\em American Journal of Public Health}, 101(2):217--223.

\bibitem[Nelson and Grossniklaus, 2019]{nelson2019trends}
Nelson, J.~M. and Grossniklaus, D.~A. (2019).
\newblock Trends in hospital breastfeeding policies in the united states from 2009--2015: Results from the maternity practices in infant nutrition and care survey.
\newblock {\em Breastfeeding Medicine}, 14(3):165--171.

\bibitem[Olivetti and Petrongolo, 2016]{olivetti2016evolution}
Olivetti, C. and Petrongolo, B. (2016).
\newblock The evolution of gender gaps in industrialized countries.
\newblock {\em Annual review of Economics}, 8:405--434.

\bibitem[Olivetti and Petrongolo, 2017]{olivetti2017economic}
Olivetti, C. and Petrongolo, B. (2017).
\newblock The economic consequences of family policies: lessons from a century of legislation in high-income countries.
\newblock {\em Journal of Economic Perspectives}, 31(1):205--30.

\bibitem[P{\'e}rez-Escamilla et~al., 2016]{perez2016impact}
P{\'e}rez-Escamilla, R., Martinez, J.~L., and Segura-P{\'e}rez, S. (2016).
\newblock Impact of the baby-friendly hospital initiative on breastfeeding and child health outcomes: a systematic review.
\newblock {\em Maternal \& child nutrition}, 12(3):402--417.

\bibitem[Roe et~al., 1999]{roe1999there}
Roe, B., Whittington, L.~A., Fein, S.~B., and Teisl, M.~F. (1999).
\newblock Is there competition between breast-feeding and maternal employment?
\newblock {\em Demography}, 36(2):157--171.

\bibitem[Singh, 2018]{singh2018integrating}
Singh, S. (2018).
\newblock Integrating motherhood and employment: the role of breastfeeding legislation.

\bibitem[Sun and Abraham, 2020]{sun2020estimating}
Sun, L. and Abraham, S. (2020).
\newblock Estimating dynamic treatment effects in event studies with heterogeneous treatment effects.
\newblock {\em Journal of Econometrics}.

\bibitem[Wouk et~al., 2016]{wouk2016clinical}
Wouk, K., Lara-Cinisomo, S., Stuebe, A.~M., Poole, C., Petrick, J.~L., and McKenney, K.~M. (2016).
\newblock Clinical interventions to promote breastfeeding by latinas: A meta-analysis.
\newblock {\em Pediatrics}, 137(1).

\end{thebibliography}

\section{APPENDIX}

\setcounter{table}{0}
\renewcommand{\thetable}{A\arabic{table}}

\setcounter{figure}{0}
\renewcommand{\thefigure}{A\arabic{figure}}

\begin{table}[htbp]
  \centering
  \caption{Law Citations By State}
\begin{tabular}{rrrrr}
    \toprule
    \multicolumn{1}{l}{\textbf{State}} & \multicolumn{1}{l}{\textbf{Statute}} &       &       &  \\
    \midrule
    \midrule
    \multicolumn{1}{l}{\textbf{Arkansas }} & \multicolumn{1}{l}{Ark Code Ann §11–5-116 (2009)} &       &       &  \\
    \multicolumn{1}{l}{\textbf{California}} & \multicolumn{1}{l}{Cal [Labor] Code §§1030–1033 (2001)} &       &       &  \\
    \multicolumn{1}{l}{\textbf{Colorado }} & \multicolumn{1}{l}{Colo Rev State §§ 8–13.5–101-104 (2008)} &       &       &  \\
    \multicolumn{1}{l}{\textbf{Connecticut }} & \multicolumn{1}{l}{Conn Gen Stat §31–40w (2001)} &       &       &  \\
    \multicolumn{1}{l}{\textbf{District of Columbia}} & \multicolumn{1}{l}{DC Code §2–1401.05 (2007)} &       &       &  \\
          & \multicolumn{1}{l}{DC Code §§2–1402.81–83 (2007)} &       &       &  \\
    \multicolumn{1}{l}{\textbf{Delaware}} & \multicolumn{1}{l}{Del. Code Ann. tit. 19, § 710] (2013)} &       &       &  \\
\bottomrule
    \end{tabular}%
    
\end{table}%

{
\def\sym#1{\ifmmode^{#1}\else\(^{#1}\)\fi}
\begin{table}[htbp]
  \centering
\caption{ACS Main Results: Labor Force Participation}
\begin{adjustbox}{width=0.85\textwidth}
\begin{tabular}{l*{6}{c}}
\hline\hline
            &\multicolumn{1}{c}{\textbf{(1)}}&\multicolumn{1}{c}{(2)}&\multicolumn{1}{c}{(3)}&\multicolumn{1}{c}{(4)}&\multicolumn{1}{c}{\textbf{(5)}}&\multicolumn{1}{c}{(6)}\\
            &\multicolumn{1}{c}{2WFE}&\multicolumn{1}{c}{2WFE}&\multicolumn{1}{c}{2WFE}&\multicolumn{1}{c}{IWES}&\multicolumn{1}{c}{IWES}&\multicolumn{1}{c}{IWES}\\
\hline
$\ell = -4$         &    -0.00641         &    -0.00641         &    -0.00641         &      -0.00972               &   -0.00972                   &  -0.00972                    \\
            &   (0.00990)         &   (0.00850)         &   (0.00983)         &       (0.01052)            &    (0.00651)                 &       (0.0047)              \\
$\ell = -3$         &     0.00729         &     0.00729         &     0.00729         &      0.00710               &       0.00710              &        0.00710\sym{*}            \\
            &    (0.0121)         &    (0.0125)         &    (0.0112)         &      (0.0097)            &  (0.0056)                   &        (0.0041)            \\
$\ell = -2$        &    -0.00674         &    -0.00674         &    -0.00674         &       -0.0080              &       -0.008\sym{*}              &          -0.008\sym{*}           \\
            &    (0.0109)         &   (0.00960)         &   (0.00899)         &       (0.00805)           &   (0.0045)                  &            (0.0041)         \\
\end{tabular}
\label{tab:ACS_results}
\end{adjustbox}
\end{table}
}

{
\def\sym#1{\ifmmode^{#1}\else\(^{#1}\)\fi}
\begin{table}[htbp]
  \centering
\caption{PSID Main Results: Labor Force Participation}
\begin{adjustbox}{width=0.85\textwidth}
\begin{tabular}{l*{4}{c}}
\hline\hline
            &\multicolumn{1}{c}{\textbf{(1)}}&\multicolumn{1}{c}{(2)}&\multicolumn{1}{c}{(3)}&\multicolumn{1}{c}{(4)}\\
           
\hline
& 2WFE & IWES & 2WFE & 2WFE\\
\hline
$\ell = -8$  & -0.019 & -0.027 & -0.019 & -0.019 \\
      & (0.044) & (0.020) & (0.050) & (0.023) \\
$\ell = -6$  & -0.033 & -0.033 & -0.033 & -0.033* \\
      & (0.038) & (0.032) & (0.040) & (0.019) \\
$\ell = -4$  & 0.011 & 0.022 & 0.011 & 0.011 \\
      & (0.032) & (0.028) & (0.052) & (0.011) \\
\end{tabular}
\label{tab:PSID_results}
\end{adjustbox}
\end{table}
}

{
\def\sym#1{\ifmmode^{#1}\else\(^{#1}\)\fi}
\begin{table}
\caption{PSID Panel Results: FLFP and Young Children}
\begin{adjustbox}{width=0.65\textwidth}
\begin{tabular}{l cc | cc}
\hline\hline
            Sample: &\multicolumn{2}{c}{\textbf{All Mothers}}&\multicolumn{2}{c}{\textbf{Mothers with $\leq 3$ Children}}\\
            Child Treatment: &\multicolumn{1}{c}{Under 3}&\multicolumn{1}{c}{Under 2}&\multicolumn{1}{c}{Under 3}&\multicolumn{1}{c}{Under 2}\\
             Control: &\multicolumn{1}{c}{Age 3-6}&\multicolumn{1}{c}{Age 2-8}&\multicolumn{1}{c}{Age 3-6}&\multicolumn{1}{c}{Age 2-8}\\
\hline
-6       &     -0.0109         &     -0.0185         &     -0.0103         &     -0.0272\sym{*}  \\            &    (0.0181)         &    (0.0136)         &    (0.0197)         &    (0.0149)         \\
-6*Child&     -0.0260         &    -0.00138         &     -0.0428\sym{*}  &    -0.00548         \\            &    (0.0225)         &    (0.0214)         &    (0.0237)         &    (0.0219)         \\
-4 &     -0.0470\sym{***}&     -0.0193\sym{*}  &    -0.00212         &    -0.00574         \\
\end{tabular}
\label{tab:PSID_interactedpanel}
\end{adjustbox}
\end{table}
}

{
\def\sym#1{\ifmmode^{#1}\else\(^{#1}\)\fi}
\begin{table}
\caption{PSID Panel Results: FLFP and Young Children, other coefficients}
\begin{adjustbox}{width=\textwidth}
\begin{tabular}{l*{4}{c}}
\hline\hline
            Sample: &\multicolumn{2}{c}{\textbf{All Mothers}}&\multicolumn{2}{c}{\textbf{Mothers with $\leq 3$ Children}}\\
            Child Treatment: &\multicolumn{1}{c}{Under 3}&\multicolumn{1}{c}{Under 2}&\multicolumn{1}{c}{Under 3}&\multicolumn{1}{c}{Under 2}\\
             Control: &\multicolumn{1}{c}{Age 3-6}&\multicolumn{1}{c}{Age 2-8}&\multicolumn{1}{c}{Age 3-6}&\multicolumn{1}{c}{Age 2-8}\\
\hline
Unmarried      &      0.0397\sym{***}&      0.0428\sym{***}&      0.0371\sym{***}&      0.0409\sym{***}\\
            &      (6.52)         &      (7.97)         &      (5.34)         &      (6.85)         \\
Education (years)    &      0.0212\sym{***}&      0.0177\sym{***}&      0.0210\sym{***}&      0.0183\sym{***}\\
            &     (10.32)         &      (9.69)         &      (8.93)         &      (8.99)         \\
Age     &    -0.00481\sym{***}&    -0.00475\sym{***}&    -0.00300\sym{***}&    -0.00318\sym{***}\\
\end{tabular}
\label{tab:PSID_interactedpanel_others}
\end{adjustbox}
\end{table}
}

{
\def\sym#1{\ifmmode^{#1}\else\(^{#1}\)\fi}
\begin{table}
\caption{FLFP: Women who Ever Breastfed (CAH Supplement)}
  \begin{adjustbox}{width=0.95\textwidth}
\begin{tabular}{l*{6}{c}}
\hline\hline
            &\multicolumn{1}{c}{(1)}&\multicolumn{1}{c}{(2)}&\multicolumn{1}{c}{(3)}&\multicolumn{1}{c}{(4)}&\multicolumn{1}{c}{(5)}&\multicolumn{1}{c}{(6)}\\
            &\multicolumn{1}{c}{2WFE}&\multicolumn{1}{c}{IWES}&\multicolumn{1}{c}{2WFE}&\multicolumn{1}{c}{IWES}&\multicolumn{1}{c}{2WFE}&\multicolumn{1}{c}{2WFE}\\
\hline
t = -6         &     -0.0241         &       -.0307           &      -0.0363         &       0.0173              &                     &                     \\
            &    (0.0633)         &      (0.0556)               &    (0.107)        &     (0.0698)               &                     &                     \\
t = -4        &       0.103\sym{***}&       0.1169\sym{***}        &      0.104        &        0.0814             &                     &                     \\
            &    (0.0291)         &     (0.0242)              &    (0.119)          &          (0.0564)           &                     &                     \\
t = 0          &      0.0972         &       0.1038\sym{**}             &      0.168\sym{*}        &         -0.0037            &                     &                     \\            &    (0.0584)         &        (0.0462)         &    (0.1000)         &        (0.0516)             &                     &                     \\
\end{tabular}
\label{tab:cah_full}
\end{adjustbox}
\end{table}
}

\begin{table}[htbp]\centering
\def\sym#1{\ifmmode^{#1}\else\(^{#1}\)\fi}
\caption{IFPS Results: Other Coefficients}
\begin{adjustbox}{width=0.6\textwidth}
\begin{tabular}{l*{5}{c}}
\hline\hline
            &\multicolumn{1}{c}{(1)}         &\multicolumn{1}{c}{(2)}         &\multicolumn{1}{c}{(3)}         &\multicolumn{1}{c}{(4)}         &\multicolumn{1}{c}{(5)}         \\
\hline
Breastfed     &    -0.00509\sym{**} &    -0.00509\sym{**} &    -0.00427\sym{**}                 &    -0.00178         &    -0.00813         \\
            &   (0.00252)         &   (0.00252)         &      (0.00187)                &   (0.00184)         &   (0.00603)         \\
[1em]
Law       &                     &      0.0217\sym{***}&       0.0216\sym{***}              &                     &                     \\
            &                     &   (0.00283)         &      (0.00318)               &                     &                     \\
[1em]
Breastfed*Law &                     &                     &    0.000560         &                     &                     \\
            &                     &                     &   (0.00346)         &                     &                     \\
[1em]
Black  &    0.000716         &    0.000716         &    0.000645         &                     &                     \\
\end{tabular}
\end{adjustbox}
\end{table}

\begin{table}[htbp]\centering
\def\sym#1{\ifmmode^{#1}\else\(^{#1}\)\fi}
\caption{Heterogeneity By Race/Ethnicity}
\begin{adjustbox}{width=0.8\textwidth}
\begin{tabular}{l c c c c | c c c}
\hline\hline 
&\multicolumn{4}{c}{\textbf{ACS Data}}         & Time & \multicolumn{2}{c}{\textbf{PSID data}}        \\
Time
            &\multicolumn{1}{c}{(White)}         &\multicolumn{1}{c}{(Black)}         &\multicolumn{1}{c}{(Non-Hispanic)}         &\multicolumn{1}{c}{(Hispanic)}     &     &\multicolumn{1}{c}{(White)}         &\multicolumn{1}{c}{(Black)}         \\
\hline
-4         &     -0.0105         &    -0.00202         &     -0.0113         &      0.0106   & -8      &      0.0689         &     -0.0695         \\
            &   (0.00849)         &    (0.0268)         &   (0.00888)         &    (0.0295)     &    &    (0.0619)         &    (0.0678)         \\
[1em]
-3         &     0.00594         &      0.0211         &     0.00813         &     -0.0120    & -6     &      0.0407         &      -0.130\sym{***}\\
            &    (0.0114)         &    (0.0368)         &    (0.0119)         &    (0.0359)  &       &    (0.0545)         &    (0.0460)         \\
[1em]
-2         &    -0.00262         &     -0.0180         &    -0.00714         &    -0.00879    &   -4   &      0.0925\sym{*}  &     -0.0702         \\
            &    (0.0104)         &    (0.0185)         &   (0.00796)         &    (0.0205)  &       &    (0.0503)         &    (0.0516)         \\
\end{tabular}
\label{tab:hetbyrace}
\end{adjustbox}
\end{table}

\begin{table}[htbp]\centering
\def\sym#1{\ifmmode^{#1}\else\(^{#1}\)\fi}
\caption{ROBUSTNESS: High vs. Low LFP Terciles}
\begin{adjustbox}{width=\textwidth}
\begin{tabular}{l*{4}{c}}
\hline\hline
       Time     &\multicolumn{1}{c}{(ACS: Tercile 1)}         &\multicolumn{1}{c}{(ACS: Tercile 3)}         &\multicolumn{1}{c}{(PSID: Tercile 1)}         &\multicolumn{1}{c}{(PSID: Tercile 3)}         \\
\hline

-4         &      0.0133         &     -0.0246         &     -0.0562         &     -0.0250         \\
            &    (0.0118)         &    (0.0263)         &    (0.0679)         &    (0.0370)         \\
[1em]
-3        &      0.0240         &     -0.0345\sym{**} &      0.0763         &     -0.0141         \\
            &    (0.0239)         &    (0.0168)         &    (0.0676)         &     (0.103)         \\
[1em]
-2        &     0.00336         &     -0.0348\sym{*}  &     -0.0312         &    0.000540         \\
            &   (0.00586)         &    (0.0200)         &    (0.0503)         &    (0.0492)         \\
[1em]
0          &      0.0257\sym{***}&     -0.0380         &      0.0554         &      0.0291         \\
\end{tabular}
\label{tab:highvlow_LFP}
\end{adjustbox}
\end{table}


\begin{table}[htbp]\centering
\def\sym#1{\ifmmode^{#1}\else\(^{#1}\)\fi}
\caption{Placebo Tests: PSID}
\begin{adjustbox}{width=\textwidth}
\begin{tabular}{l*{8}{c}}
\hline\hline
         Time   &\multicolumn{1}{c}{(Age $>45$)}         &\multicolumn{1}{c}{(Age $>45$)}         &\multicolumn{1}{c}{(Child age $>6$)}         &\multicolumn{1}{c}{(Child age $>6$)}         &\multicolumn{1}{c}{(No kids)}         &\multicolumn{1}{c}{(No kids)}         &\multicolumn{1}{c}{(Never kids)}         &\multicolumn{1}{c}{(Never kids)}         \\
\hline
-8         &    -0.00605         &      0.0204\sym{*}  &     -0.0106         &     0.00483         &     -0.0122         &      0.0127         &    -0.00487         &    -0.00289         \\
            &    (0.0212)         &    (0.0117)         &    (0.0137)         &   (0.00762)         &    (0.0134)         &    (0.0119)         &    (0.0208)         &    (0.0171)         \\
[1em]
-6        &    -0.00588         &     0.00573         &     0.00504         &     0.00124         &     -0.0361\sym{**} &     -0.0227\sym{**} &     -0.0336         &     -0.0410\sym{***}\\
            &    (0.0245)         &    (0.0110)         &    (0.0182)         &   (0.00730)         &    (0.0143)         &    (0.0108)         &    (0.0240)         &    (0.0155)         \\
[1em]
-4         &    -0.00872         &     0.00388         &    0.000697         &    -0.00349         &     -0.0302\sym{***}&     -0.0182\sym{**} &    -0.00838         &     -0.0156         \\
            &    (0.0191)         &   (0.00878)         &    (0.0102)         &   (0.00606)         &    (0.0106)         &   (0.00853)         &    (0.0168)         &    (0.0117)         \\
[1em]
0         &     0.00512         &      0.0127         &   -0.000898         &     0.00223         &     -0.0155         &     0.00239         &     0.00456         &     0.00189         \\
\end{tabular}
\end{adjustbox}
\label{tab:placebos}
\end{table}

\begin{table}[htbp]\centering
\def\sym#1{\ifmmode^{#1}\else\(^{#1}\)\fi}
\caption{ROBUSTNESS: Other Labor Policies}
\begin{adjustbox}{width=0.8\textwidth}
    \begin{tabular}{l c c c | c c c }
\hline\hline
& \multicolumn{3}{c}{\textbf[ACS]} & \multicolumn{3}{c}{\textbf[PSID]} \\
        Time    &\multicolumn{1}{c}{(1)}         &\multicolumn{1}{c}{(2)}         &\multicolumn{1}{c}{(3)}         &\multicolumn{1}{c}{(4)}         &\multicolumn{1}{c}{(5)}         &\multicolumn{1}{c}{(6)}         \\
\hline

-4         &    -0.00625         &    -0.00639         &    -0.00624         &     -0.0250         &     -0.0230         &     -0.0307         \\
            &   (0.00850)         &   (0.00849)         &   (0.00850)         &    (0.0424)         &    (0.0445)         &    (0.0428)         \\
[1em]
-3         &     0.00742         &     0.00731         &     0.00744         &     -0.0388         &     -0.0345         &     -0.0417         \\
            &    (0.0124)         &    (0.0125)         &    (0.0124)         &    (0.0351)         &    (0.0384)         &    (0.0354)         \\
[1em]
-2         &    -0.00671         &    -0.00673         &    -0.00671         &      0.0155         &     0.00834         &      0.0135         \\
            &   (0.00957)         &   (0.00959)         &   (0.00956)         &    (0.0307)         &    (0.0320)         &    (0.0310)         \\
\end{tabular}
\label{tab:otherpolicies}
\end{adjustbox}
\end{table}

\begin{table}[htbp]\centering
\def\sym#1{\ifmmode^{#1}\else\(^{#1}\)\fi}
\caption{Leave One State Out, Page 1}
\begin{adjustbox}{width=\textwidth}
    \begin{tabular}{l*{10}{c}}
\hline\hline
      Without:      &\multicolumn{1}{c}{(Alabama)}         &\multicolumn{1}{c}{(Alaska)}         &\multicolumn{1}{c}{(Arizona)}         &\multicolumn{1}{c}{(Arkansas)}         &\multicolumn{1}{c}{(Colorado)}         &\multicolumn{1}{c}{(Delaware)}         &\multicolumn{1}{c}{(DC)}         &\multicolumn{1}{c}{(Florida)}         &\multicolumn{1}{c}{(Hawaii)}         &\multicolumn{1}{c}{(Idaho)}         \\
\hline

-4         &    -0.00626         &    -0.00656         &    -0.00693         &    -0.00449         &    -0.00910         &    -0.00756         &    -0.00617         &    -0.00442         &     -0.0101         &    -0.00660         \\
            &   (0.00852)         &   (0.00855)         &   (0.00852)         &   (0.00907)         &   (0.00809)         &   (0.00851)         &   (0.00851)         &   (0.00814)         &   (0.00773)         &   (0.00850)         \\
[1em]
-3         &     0.00727         &     0.00712         &     0.00717         &     0.00708         &     0.00325         &     0.00576         &     0.00749         &     0.00886         &     0.00614         &     0.00702         \\
            &    (0.0124)         &    (0.0124)         &    (0.0125)         &    (0.0133)         &    (0.0135)         &    (0.0125)         &    (0.0125)         &    (0.0125)         &    (0.0128)         &    (0.0124)         \\
[1em]
-2         &    -0.00681         &    -0.00687         &    -0.00680         &    -0.00801         &     -0.0153\sym{*} &    -0.00630         &    -0.00689         &    -0.00587         &    -0.00590         &    -0.00691         \\
            &   (0.00962)         &   (0.00961)         &   (0.00967)         &   (0.00994)         &   (0.00573)         &   (0.00973)         &   (0.00970)         &   (0.00945)         &   (0.00980)         &   (0.00963)         \\
\end{tabular}
\end{adjustbox}

\end{table}

\begin{table}[htbp]\centering
\def\sym#1{\ifmmode^{#1}\else\(^{#1}\)\fi}
\caption{Leave One State Out, Page 2}
\begin{adjustbox}{width=\textwidth}
    \begin{tabular}{l*{10}{c}}
\hline\hline
      Without:      &\multicolumn{1}{c}{(Indiana)}         &\multicolumn{1}{c}{(Iowa)}         &\multicolumn{1}{c}{(Kansas)}         &\multicolumn{1}{c}{(Kentucky)}         &\multicolumn{1}{c}{(Louisiana)}         &\multicolumn{1}{c}{(Maine)}         &\multicolumn{1}{c}{(Maryland)}         &\multicolumn{1}{c}{(Mass)}         &\multicolumn{1}{c}{(Michigan)}         &\multicolumn{1}{c}{(Miss)}         \\
\hline

-4         &    -0.00552         &    -0.00654         &    -0.00618         &    -0.00646         &    -0.00627         &    -0.00463         &    -0.00705         &    -0.00883         &    -0.00628         &    -0.00628         \\
            &   (0.00843)         &   (0.00854)         &   (0.00853)         &   (0.00848)         &   (0.00856)         &   (0.00856)         &   (0.00847)         &   (0.00919)         &   (0.00848)         &   (0.00858)         \\
[1em]
-3         &     0.00786         &     0.00726         &     0.00748         &     0.00726         &     0.00738         &     0.00842         &     0.00687         &      0.0144         &     0.00732         &     0.00744         \\
            &    (0.0125)         &    (0.0125)         &    (0.0124)         &    (0.0125)         &    (0.0125)         &    (0.0128)         &    (0.0125)         &    (0.0122)         &    (0.0126)         &    (0.0125)         \\
[1em]
-2         &    -0.00665         &    -0.00674         &    -0.00664         &    -0.00676         &    -0.00644         &    -0.00524         &    -0.00669         &    -0.00691         &    -0.00681         &    -0.00641         \\
            &   (0.00966)         &   (0.00964)         &   (0.00959)         &   (0.00959)         &   (0.00957)         &   (0.00964)         &   (0.00943)         &    (0.0105)         &   (0.00964)         &   (0.00951)         \\
\end{tabular}
\end{adjustbox}

\end{table}

\begin{table}[htbp]\centering
\def\sym#1{\ifmmode^{#1}\else\(^{#1}\)\fi}
\caption{Leave One State Out, Page 3}
\begin{adjustbox}{width=\textwidth}
    \begin{tabular}{l*{10}{c}}
\hline\hline
      Without:      &\multicolumn{1}{c}{(Missouri)}         &\multicolumn{1}{c}{(Montana)}         &\multicolumn{1}{c}{(Nebraska)}         &\multicolumn{1}{c}{(Nevada)}         &\multicolumn{1}{c}{(NH)}         &\multicolumn{1}{c}{(New Jersey)}         &\multicolumn{1}{c}{(New Mexico)}         &\multicolumn{1}{c}{(New York)}         &\multicolumn{1}{c}{(NC)}         &\multicolumn{1}{c}{(ND)}         \\
\hline

-4         &    -0.00688         &    -0.00512         &    -0.00268         &    -0.00654         &    -0.00637         &    -0.00727         &    -0.00663         &    -0.00486         &    -0.00622         &    -0.00640         \\
            &   (0.00862)         &   (0.00843)         &   (0.00813)         &   (0.00894)         &   (0.00847)         &   (0.00849)         &   (0.00851)         &   (0.00899)         &   (0.00874)         &   (0.00848)         \\
[1em]
-3         &     0.00698         &     0.00865         &     0.00815         &     0.00959         &     0.00732         &     0.00682         &     0.00709         &     0.00849         &     0.00716         &     0.00723         \\
            &    (0.0124)         &    (0.0125)         &    (0.0130)         &    (0.0130)         &    (0.0125)         &    (0.0124)         &    (0.0125)         &    (0.0128)         &    (0.0122)         &    (0.0125)         \\
[1em]
-2         &    -0.00709         &    -0.00582         &    -0.00389         &    -0.00451         &    -0.00670         &    -0.00695         &    -0.00718         &    -0.00316         &    -0.00668         &    -0.00681         \\
            &   (0.00973)         &   (0.00969)         &   (0.00942)         &   (0.00961)         &   (0.00959)         &   (0.00958)         &   (0.00978)         &    (0.0115)         &   (0.00960)         &   (0.00962)         \\
\end{tabular}
\end{adjustbox}

\end{table}

\begin{table}[htbp]\centering
\def\sym#1{\ifmmode^{#1}\else\(^{#1}\)\fi}
\caption{Leave One State Out, Page 4}
\begin{adjustbox}{width=\textwidth}
    \begin{tabular}{l*{12}{c}}
\hline\hline
      Without:      &\multicolumn{1}{c}{(Ohio)}         &\multicolumn{1}{c}{(OK)}         &\multicolumn{1}{c}{(Oregon)}         &\multicolumn{1}{c}{(PA)}         &\multicolumn{1}{c}{(SC)}         &\multicolumn{1}{c}{(SD)}         &\multicolumn{1}{c}{(Texas)}         &\multicolumn{1}{c}{(Utah)}         &\multicolumn{1}{c}{(Virginia)}         &\multicolumn{1}{c}{(Washington)}  &\multicolumn{1}{c}{(WV)} &\multicolumn{1}{c}{(WI)}        \\
\hline

-4         &    -0.00671         &    -0.00825         &    -0.00695         &    -0.00642         &    -0.00568         &    -0.00627         &    -0.00793         &    -0.00781         &    -0.00156         &    -0.00732         &    -0.00600         &    -0.00659         \\
            &   (0.00876)         &   (0.00837)         &   (0.00851)         &   (0.00847)         &   (0.00945)         &   (0.00847)         &   (0.00868)         &   (0.00897)         &   (0.00899)         &   (0.00975)         &   (0.00848)         &   (0.00851)         \\
[1em]
-3         &     0.00705         &     0.00549         &     0.00658         &     0.00725         &     0.00979         &     0.00749         &     0.00604         &    0.000599         &     0.00211         &      0.0136         &     0.00760         &     0.00701         \\
            &    (0.0123)         &    (0.0124)         &    (0.0124)         &    (0.0127)         &    (0.0138)         &    (0.0125)         &    (0.0122)         &    (0.0117)         &    (0.0133)         &    (0.0127)         &    (0.0124)         &    (0.0125)         \\
[1em]
-2         &    -0.00708         &    -0.00847         &    -0.00771         &    -0.00676         &    -0.00478         &    -0.00653         &    -0.00743         &    -0.00625         &    -0.00752         &    -0.00652         &    -0.00649         &    -0.00703         \\
\end{tabular}
\end{adjustbox}

\end{table}

\begin{figure}[H]
\centering
\begin{minipage}{0.65\textwidth} 
\caption{Laws For Public Employees}
\includegraphics[width=10cm]{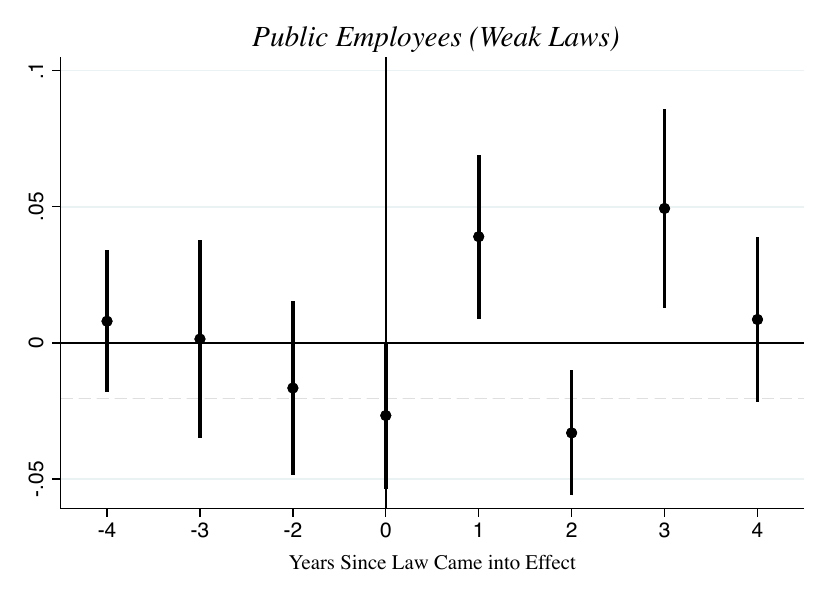}
{\footnotesize \justifying \singlespacing{This figure depicts results for 2WFE regressions with the treatment re-defined as the laws only requiring provision of workplace amenities for public employees, with 95\% confidence intervals using the ACS 2005-2019 main sample, which consists of women who gave birth in the last year. The control group is only states which never pass the law, and those with stronger laws targeting the entire labor force are removed from the analysis. Standard errors are clustered at the state level. } \par}
\end{minipage}
\label{fig:publicemps}
\end{figure}

\end{document}